\documentclass[twocolumn]{aastex63}
\usepackage{newtxtext,newtxmath}
\usepackage[utf8]{inputenc}
\usepackage{amsmath,amssymb}

\usepackage{graphicx}
\usepackage[caption=false]{subfig}
\received{\today}
\revised{\today}
\accepted{\today}
\submitjournal{ApJS}

\shorttitle{GRRMHD with robust primitive variable recovery}
\shortauthors{B. Ripperda, F. Bacchini et al.}
\begin{document}

\title{General relativistic resistive magnetohydrodynamics with robust primitive variable recovery for accretion disk simulations}
\correspondingauthor{B. Ripperda}\email{ripperda@itp.uni-frankfurt.de}
\correspondingauthor{F. Bacchini}\email{bacchini@itp.uni-frankfurt.de}
\author[0000-0002-7301-3908]{B. Ripperda}\altaffiliation{Both first authors have contributed equally to all aspects of this work.}\affiliation{Institut f\"{u}r Theoretische Physik, Goethe-Universit\"{a}t, Max-von-Laue-Str. 1, D-60438 Frankfurt am Main, Germany}\affiliation{Centre for mathematical Plasma-Astrophysics, Department of Mathematics, Katholieke Universiteit Leuven, Celestijnenlaan 200B, B-3001 Leuven, Belgium}
\author[0000-0002-0281-2745]{F. Bacchini}\affiliation{Institut f\"{u}r Theoretische Physik, Goethe-Universit\"{a}t, Max-von-Laue-Str. 1, D-60438 Frankfurt am Main, Germany}\affiliation{Centre for mathematical Plasma-Astrophysics, Department of Mathematics, Katholieke Universiteit Leuven, Celestijnenlaan 200B, B-3001 Leuven, Belgium}
\author[0000-0002-4584-2557]{O. Porth}\affiliation{Astronomical Institute Anton Pannekoek, Universeit van Amsterdam, Science Park 904, 1098 XH, Amsterdam, The Netherlands}\affiliation{Institut f\"{u}r Theoretische Physik, Goethe-Universit\"{a}t, Max-von-Laue-Str. 1, D-60438 Frankfurt am Main, Germany}
\author[0000-0002-0491-1210]{E. R. Most}\affiliation{Institut f\"{u}r Theoretische Physik, Goethe-Universit\"{a}t, Max-von-Laue-Str. 1, D-60438 Frankfurt am Main, Germany}
\author[0000-0001-6833-7580]{H. Olivares}\affiliation{Institut f\"{u}r Theoretische Physik, Goethe-Universit\"{a}t, Max-von-Laue-Str. 1, D-60438 Frankfurt am Main, Germany}
\author{A. Nathanail}\affiliation{Institut f\"{u}r Theoretische Physik, Goethe-Universit\"{a}t, Max-von-Laue-Str. 1, D-60438 Frankfurt am Main, Germany}
\author[0000-0002-1330-7103]{L. Rezzolla}\affiliation{Institut f\"{u}r Theoretische Physik, Goethe-Universit\"{a}t, Max-von-Laue-Str. 1, D-60438 Frankfurt am Main, Germany}
\author[0000-0003-0811-5091]{J. Teunissen}\affiliation{Centre for mathematical Plasma-Astrophysics, Department of Mathematics, Katholieke Universiteit Leuven, Celestijnenlaan 200B, B-3001 Leuven, Belgium}\affiliation{Centrum Wiskunde \& Informatica, Amsterdam, The Netherlands}
\author[0000-0003-3544-2733]{R. Keppens}\affiliation{Centre for mathematical Plasma-Astrophysics, Department of Mathematics, Katholieke Universiteit Leuven, Celestijnenlaan 200B, B-3001 Leuven, Belgium}


\begin{abstract}
Recent advances in black
  hole astrophysics, particularly the first visual evidence of a
  supermassive black hole at the center of the galaxy M87 by the Event
  Horizon Telescope (EHT), and the detection of an orbiting ``hot spot''
  nearby the event horizon of Sgr A$\mbox{*}$ in the Galactic center by
  the Gravity Collaboration, require the development of novel numerical
  methods to understand the underlying plasma microphysics. Non-thermal
  emission related to such hot spots is conjectured to originate from
  plasmoids that form due to magnetic reconnection in thin current layers
  in the innermost accretion zone. Resistivity plays a crucial role in
current sheet formation, magnetic reconnection, and plasmoid growth in
black hole accretion disks and jets. We included resistivity in the
three-dimensional general-relativistic magnetohydrodynamics (GRMHD) code
{\tt BHAC} and present the implementation of an Implicit-Explicit scheme
to treat the stiff resistive source terms of the GRMHD equations. The
algorithm is tested in combination with adaptive mesh refinement to
resolve the resistive scales and a constrained transport method to keep
the magnetic field solenoidal. Several novel methods for primitive
variable recovery, a key part in relativistic magnetohydrodynamics codes,
are presented and compared for accuracy, robustness, and efficiency. We
propose a new inversion strategy that allows for resistive-GRMHD
simulations of low gas-to-magnetic pressure ratio and highly magnetized
regimes as applicable for black hole accretion disks, jets, and neutron
star magnetospheres. We apply the new scheme to
study the effect of resistivity on accreting black holes, accounting for
dissipative effects as reconnection.
\end{abstract}




\section{Introduction}

Astrophysical phenomena typically show very distinctive time and length
scales on which microscopic and macroscopic dynamics take
place. Relativistic macroscopic plasma dynamics can be described by
general-relativistic magnetohydrodynamics (GRMHD), coupling the fluid of
charged particles to electromagnetic fields in a dynamic or static
gravitational field. This framework explains many observed
astrophysical-plasma phenomena on the global scale, such as accretion onto and outflows from
compact objects. Despite outstanding results achieved with GRMHD,
state-of-the-art studies are affected by a lack of information on the
effect of microscopic plasma physics on the macroscopic dynamics. The
magnetorotational instability (MRI), for example, is a crucial mechanism
of angular momentum transport in accretion disks resulting in turbulent
motion (\citealt{velikhov1959}; \citealt{chandrasekhar1960};
\citealt{balbus1991}). Magnetic reconnection and subsequent particle
acceleration can occur in the turbulent disk or in the highly magnetized
jet. Field-amplifying processes like the MRI, and dissipative processes
like turbulence and reconnection, typically occur across a large range of
scales from microscopic to macroscopic. Accurate modeling of the rarefied
magnetospheres of compact objects requires knowledge of such fundamental
microscopic processes affecting the macroscopic dynamics. Dynamic
electric and magnetic fields determine the spatial and temporal scales on
which dissipation occurs. Examples of astrophysical systems where both
the electric and the magnetic fields are important include: accreting
black holes in active galactic nuclei, coalescing black-hole and
neutron-star binaries, neutron-star and magnetar magnetospheres,
pulsar-wind nebulae, and massive stars undergoing core collapse. The
magnetic field can change its topology via magnetic reconnection
resulting in dissipation of the released magnetic energy that accelerates
particles causing non-thermal emission. Non-thermal emission is one of
the main uncertainties in the ideal-GRMHD (i.e.,
with infinite electrical conductivity $\sigma$) models of the Event
Horizon Telescope (EHT) observations of the accretion disk of
M87$\mbox{*}$, the supermassive black hole at the center of the galaxy
M87 (\citealt{EHTpaper1}; \citealt{EHTpaper5}). Sgr A$\mbox{*}$, the
black hole at the center of the Milky Way, also regularly exhibits
non-thermal emission in the form of flares that have been conjectured to
originate from magnetic reconnection in the accretion disk
(\citealt{baganoff2001}; \citealt{genzel2003}; \citealt{eckart2006};
\citealt{Meyer2008}; \citealt{neilsen2013}; \citealt{Dexter2014};
\citealt{brinkerink2015}; \citealt{Gravity2018}). Dissipative (non-ideal)
effects that can cause non-thermal emission are often
negligible, except in regions with a strong and localized current
density, e.g., in long and thin current sheets. Such effects occur on the
diffusion time scale $\tau_D = L^2/\eta$, where $L$ is the characteristic
length scale of the system and $\eta$ is the resistivity. Plasma is
typically collisionless in astrophysical systems like Sgr A$\mbox{*}$ and
M87$\mbox{*}$, such that diffusion time scales are much larger than
Alfv\'{e}nic time scales $\tau_A = L/v_A$, with $v_A$ the Alfv\'{e}n
speed. The gyroradius of electrons and ions can be considered as an
effective mean free path perpendicular to the magnetic field, which is
typically orders of magnitude smaller than the typical system size
$r_{\rm g} = GM/c^2$, the Schwarzschild radius with gravitational
constant $G$, mass of the object $M$ and speed of light $c$. The mean
free path along the magnetic field is typically $\gg r_{\rm g}$, such
that the particles can freely travel along magnetic field lines before
being deflected by Coulomb collisions (\citealt{yuan2014};
\citealt{porth2019}). Hence, $\tau_D \gg \tau_A$ and treating the plasma
as an ideal magnetized fluid is a reasonable approach. However, to
capture dissipation physics like magnetic reconnection, the diffusive
time scale needs to be resolved in a practically near-dissipation-less
system (i.e., $\tau_D \gg \tau_A$). Resolving such different time scales
requires specific numerical schemes that can handle both fast and slow
dynamics. Additionally, dissipative dynamics take place on a large range
of length scales, which can depend both on the explicit resistivity and
on the grid resolution in a numerical simulation. Hence, a sufficiently
high resolution is essential to capture small resistive length scales in
GRMHD simulations.

The set of ideal-MHD (i.e., $\eta=0$) equations cannot describe non-ideal
processes due to the frozen-in condition of the magnetic field
(\citealt{Alfven1942}). Ideal-GRMHD simulations have
been conducted to explore magnetic reconnection in accretion disks as
triggered by numerical resistivity (\citealt{ball2016};
\citealt{ball2017}), yet in this case there is no control on the
resistivity, which purely depends on the numerical method and resolution used,
rather than on a underlying physical model. The framework of
general-relativistic resistive magnetohydrodynamics (GRRMHD) allows for
incorporating a physical resistivity and to systematically explore
dissipative effects like magnetic reconnection. In the set of
non-relativistic magnetohydrodynamics equations, a resistive source term
for resistivity $\eta$ can be added directly to the induction equation
and the electric field depends on the magnetic field $\mathbf{B}$, the
current density $\mathbf{J} = \nabla \times \mathbf{B}$, and the fluid
velocity field $\mathbf{v}$ via a simple algebraic expression, i.e.,
$\mathbf{E}(\mathbf{v}, \mathbf{B})=-\mathbf{v}\times \mathbf{B} +\eta
\mathbf{J}$. In this way, the system remains hyperbolic-parabolic
allowing for a standard explicit integration method. To incorporate
resistivity in GRMHD however, Amp\`{e}re's law (i.e., for the evolution
of the electric field) has to be solved alongside the ideal-GRMHD
equations and one cannot assume $\mathbf{E}=-\mathbf{v}\times \mathbf{B}
+\eta \mathbf{J}$. Additionally, the induction equation in ideal-MHD
assumes the electric field to be a purely dependent variable, i.e.,
$\mathbf{E} (\mathbf{v}, \mathbf{B})=-\mathbf{v}\times \mathbf{B}$, such
that it has to be replaced by Faraday's law (i.e., for the evolution of
the magnetic field) to account for a resistive electric field in relativistic MHD. For a
realistically small but finite resistivity, the electric field dynamics
occurs on a much shorter time scale than the GRMHD evolution. This
results in a stiff source term in Amp\`{e}re's law which makes the time
evolution with an explicit time integrator very inefficient.

Many methods have been developed in recent years to handle the additional
complexity of the unavoidable stiff resistive source term in relativistic
magnetohydrodynamics. \cite{Komissarov} presented a special-relativistic
resistive magnetohydrodynamics (SRRMHD) scheme based on a Strang-split
method (\citealt{Strang}), approximating the resistive source terms
resulting from Ohm's law with a semi-analytic approach. The Strang-split
method requires a time-step that is proportional to the resistivity,
resulting in very expensive computations for typical astrophysical
plasmas with extremely low resistivity. In \cite{dumbser2009}, an
unstructured mesh approach was suggested to solve the SRRMHD
equations. \cite{Palenzuela2} presented the first application of an
implicit-explicit (IMEX) Runge-Kutta (RK) scheme of arbitrary order 
to SRRMHD simulations, incorporating the full resistive Ohm's law. There, the 
stiff source term is solved implicitly, while the non-stiff equations are solved 
explicitly as in relativistic ideal-MHD. This overcomes the limitations otherwise
imposed on the time-step due to stiffness, preventing a major slow-down
compared to relativistic ideal-MHD. \cite{takamoto2011} improved the
semi-analytic approach of \cite{Komissarov} with an implicit inversion
method for the stiff source term, relaxing the restrictive time-step for
SRRMHD. \cite{Bucciantini} and \cite{Dionysopoulou} applied the IMEX
method to GRRMHD incorporating the full resistive Ohm's law and
\cite{bodomignone} recently presented substantial improvements in
SRRMHD. \cite{Bucciantini} and \cite{Palenzuela} used an IMEX scheme to
include Hall and dynamo effects in the GRRMHD evolution, through a
generalized Ohm's law. Numerical methods for GRRMHD have been extensively
applied to neutron-star mergers (\citealt{Palenzuela2013}; \citealt{Palenzuela2013b}; \citealt{dionysopoulou2015}), the
collapse of a neutron star to a black hole (\citealt{Palenzuela};
\citealt{nathanail2017}; \citealt{most2018}), accretion onto black holes
(\citealt{bugli2014}; \citealt{Qian2016}; \citealt{qian2018}; \citealt{vourellis2019}) and in
SRRMHD for relativistic reconnection (\citealt{zenitani2010}; \citealt{barkov2013};
\citealt{mizuno2013}; \citealt{delzanna2016}; \citealt{ripperda2019}).


Numerical schemes for GRMHD require a method to recover ``primitive''
variables such as rest-mass density, pressure and the four-velocity from
a set of ``conserved'' variables such as momentum and energy density. To
retrieve the primitive variables it is necessary to solve one or more
nonlinear equations. The solution method for the nonlinear equations is
essential and is often a bottleneck for both accuracy and computational
costs (\citealt{noble2006}; \citealt{siegel2018}). For stiff systems such
as the set of GRRMHD equations, where the electric field is dynamically
important, the primitive variables depend nonlinearly on the electric
field and vice versa, resulting in an additional complication in the
primitive variable recovery compared to ideal-GRMHD. Standard
primitive-recovery methods for GRRMHD, often naively adapted from GRMHD,
do not account for the nonlinear dependence of the electric field on the
primitive variables (\citealt{Palenzuela2},
\citealt{Dionysopoulou}, \citealt{Palenzuela}; \citealt{Qian2016}), and
are therefore less robust in highly magnetized plasma regions that are
frequently encountered around black holes and neutron stars.

In this work, we implement the IMEX method of \citealt{Bucciantini}
combined with several novel and robust primitive-recovery methods for
GRRMHD in the Black Hole Accretion Code ({\tt BHAC}, \citealt{BHAC}), a
versatile general-relativistic magneto-fluid code based on the {\tt
  MPI-AMRVAC} framework (\citealt{vanderholst_2008};
\citealt{keppens_2012}; \citealt{Keppensporth}; \citealt{Xia_2017}). The
designed recovery methods fully incorporate the electric field dynamics,
such that highly magnetized regions around black holes and neutron stars
can be accurately resolved in the resistive regime in between the
electrovacuum ($\eta \rightarrow \infty$) and the ideal-MHD limits ($\eta
\rightarrow 0$). The methods are compared to the standard recovery
schemes as presented by \cite{Palenzuela2}, \cite{Bucciantini},
\cite{Dionysopoulou}, and \cite{bodomignone}. We provide full details 
of the recovery procedure, such that it can be readily implemented in 
GRRMHD algorithms. We also propose a fall-back strategy if one or more 
methods fail to retrieve the primitive variables. The various methods are 
assessed for their accuracy, computational cost, and robustness in a survey 
over different parameter spaces and in several one- and multidimensional tests 
that are relevant for astrophysics.

In addition to having to deal with small time scales, resistive
relativistic simulations have to resolve dissipative phenomena that occur
across multiple spatial scales. With a uniform mesh, the computational
costs of large-scale simulations with enough resolution to resolve the
dissipative processes rapidly becomes prohibitive. An effective solution
for problems where it is essential to simultaneously resolve microscopic
and macroscopic dynamics can be found in adaptive mesh refinement (AMR)
techniques. With these methods, the underlying grid on which the
calculations are done is refined during the simulation. Adopting criteria
that are based on the properties of the plasma dynamics, a finer grid is
introduced in order to accurately resolve smaller scales in a confined
area, thus dramatically reducing the computational costs
(\citealt{keppens2003}; \citealt{BHAC}). A constrained transport (CT)
method that is compatible with AMR is employed to keep the divergence of
the magnetic field equal to machine precision at all times
(\citealt{olivares2018proc}; \citealt{Olivares2019}). The algorithm is
developed to solve the GRRMHD equations in any spacetime metric in either
one, two, or three spatial dimensions.

The paper is organized as follows: Sec. \ref{sect:theory} contains the
GRRMHD equations and illustrates the main differences with the special
relativistic and non-relativistic limits. Section
\ref{sect:numericalgrrmhd} describes the numerical methods that are used
to solve the GRRMHD equations. In Sec. \ref{sect:tests} these methods
are tested for well-known cases in special and general-relativistic
magnetohydrodynamics and the accuracy of different methods for the
conserved to primitive variable transformation is explored. Our findings
are summarized in Sec. \ref{sect:conclusions}.

\section{General-Relativistic Resistive Magnetohydrodynamics}
\label{sect:theory}

In this section we briefly describe the covariant GRRMHD equations and
introduce the notation as used in this paper. We mainly emphasize the
differences between GRRMHD and the ideal-GRMHD equations solved in
\texttt{BHAC}. More information and details on the numerical schemes and
on the form of the chosen equations can be found in \cite{BHAC}. We
follow the derivation of the GRRMHD equations as in
\cite{Bucciantini}. For the remainder of this paper, we choose a
$(-,+,+,+)$ signature for the spacetime metric. Units are adopted in
which the speed of light, $c=1$, vacuum permeability $\mu_0 = 1$, vacuum
permittivity $\epsilon_0 = 1$, the gravitational constant $G=1$, and all
factors $4\pi=1$. When considering curved spacetimes, all masses are
normalized to the mass of the central object. Greek indices run over
space and time, i.e., (0,1,2,3), and Roman indices run over space only
i.e., (1,2,3).

\subsection{$3+1$ formulation of general relativity}
In the context of numerically solving the GRRMHD equations, it is useful
to write the equations in the $3+1$ form based on the
Arnowitt-Deser-Misner (ADM) formalism (see e.g.,
\citealt{RezollaBook}). We introduce the foliation of space-like
hypersurfaces $\Sigma_t$, defined as iso-surfaces of a scalar time
function $t$, and a time-like unit vector that is normal to these
hypersurfaces (\citealt{BHAC})
\begin{equation}
n_{\mu} := -\alpha \nabla_{\mu}t,
\label{eq:unitvec}
\end{equation}
where $\alpha$ is the lapse function. The frame of the \textit{Eulerian
  observer} is defined by the four-velocity $n^{\mu}$ and the metric
associated with each slice $\Sigma_t$ can be written as
\begin{equation}
\gamma_{\mu\nu} := g_{\mu\nu} + n_{\mu}n_{\nu}.
\label{eq:timeslicemetric}
\end{equation}
The spatial projection operator is then chosen
\begin{equation}
{\gamma^{\mu}}_{\nu} :={\delta^{\mu}}_{\nu} + n^{\mu}n_{\nu},
\label{eq:spatialprojection}
\end{equation}
thus satisfying the constraint ${\gamma^{\mu}}_{\nu}n_{\mu} = 0$. This
can be used to project any four-vector or tensor into its spatial and
temporal component. In this formulation, any metric can be written in the
form
\begin{equation}
 g_{\mu\nu}=
 \begin{pmatrix}
  -\alpha^2+\beta_k\beta^k & \beta_j \\
  \beta_j & \gamma_{ij}
 \end{pmatrix},
 \label{eq:3p1metric}
\end{equation}
where $\beta^i$ is the shift three-vector, and $\gamma_{ij}$ is the
three-metric representing the spatial part of $g_{\mu\nu}$, with
determinant $\gamma$ for which $(-g)^{1/2} := \alpha \gamma^{1/2}$. The
corresponding inverse metric reads
\begin{equation}
 g^{\mu\nu}=
 \begin{pmatrix}
  -1/\alpha^2 &\beta^j/\alpha^2 \\
  \beta^j/\alpha^2 & \gamma^{ij}-\beta^i\beta^j/\alpha^2
 \end{pmatrix},
 \label{eq:3p1invmetric}
\end{equation}
where $\gamma^{ij}$ is the algebraic inverse of $\gamma_{ij}$, and
$\beta^i=\gamma^{ij}\beta_j$.  It is generally straightforward to obtain
the expressions of $\alpha$, $\beta^i$ and $\gamma_{ij}$ from the
standard formulation of any general-relativistic metric (see
\citealt{BHAC} for commonly used metrics in {\tt BHAC}). Special
relativity is trivially retrieved by setting $\alpha = 1$, $\beta^i = 0$,
and $\gamma^{ij}=\delta^{ij}$. In the $3+1$ formalism, the line element is
written
\begin{equation}
ds^2 = -\alpha^2dt^2 +\gamma_{ij}\left(dx^i +\beta^i dt\right) \left(dx^j + \beta^j dt\right),
\label{eq:linelement}
\end{equation}
describing the motion of coordinate lines as seen by an Eulerian observer
\begin{equation}
x^i_{t+dt} = x^i_t - \beta^i(t,x^j)dt,
\label{eq:coordinates}
\end{equation}
moving with four-velocity
\begin{equation}
n_{\mu} = (-\alpha,0,0,0), \qquad n^{\mu} = (1/\alpha, -\beta^i/\alpha).
\label{eq:fourvelocityeulerian}
\end{equation}
A fluid element with four-velocity $u^{\mu}$ has a Lorentz factor $\Gamma
:= - u^{\mu}n_{\mu} = \alpha u^0 = (1-v^2)^{-1/2}$ with $v^2 := v_i
v^i$. This defines the fluid three-velocity
\begin{equation}
v^i := \frac{{\gamma^i}_{\mu}u^{\mu}}{\Gamma} = \frac{u^i}{\Gamma} + \frac{\beta^i}{\alpha}, \qquad v_i := \gamma_{ij}v^j = \frac{u_i}{\Gamma}.
\label{eq:fluidvelocity}
\end{equation}

\subsection{The  fluid conservation equations}
The fluid equations in general relativity are written as a set of
conservation laws for mass
\begin{equation}
\nabla_{\mu} \left(\rho u^{\mu} \right) = 0,
\label{eq:mass}
\end{equation}
and energy and momentum
\begin{equation}
\nabla_{\mu} T^{\mu \nu} = 0,
\label{eq:energy}
\end{equation}
where $\rho$ is the rest-mass density. The stress-energy tensor for a
magnetized perfect fluid is written (\citealt{Dionysopoulou};
\citealt{Qian2016})
\begin{equation}
T^{\mu \nu} \equiv T^{\mu\nu}_\mathrm{fluid} +T^{\mu\nu}_\mathrm{EM},
\label{eq:stressenergy}
\end{equation}
where the fluid part is expressed independently of the electromagnetic
fields (see e.g., \citealt{gammie2003}):
\begin{equation}
T^{\mu\nu}_\mathrm{fluid} = \left[\rho\left(1 + \epsilon\right) + p\right]u^{\mu} u^{\nu} + p g^{\mu \nu},
\label{eq:stressenergyfluid}
\end{equation} 
with fluid pressure $p$ and specific internal energy $\epsilon$. The
electromagnetic part is generally given by
\begin{equation}
 T^{\mu\nu}_\mathrm{EM} = F^{\mu\alpha}{F^\nu}_\alpha - \frac{1}{4}g^{\mu\nu}F_{\alpha\beta}F^{\alpha\beta},
\label{eq:Tempart}
\end{equation}
where $F^{\mu\nu}$ is the Maxwell tensor with Hodge dual
${}^*F^{\mu\nu}$, the Faraday tensor.

Applying the $3+1$ split and assuming a stationary spacetime, the
conservation Eqs. \eqref{eq:mass}--\eqref{eq:energy} can be written in
the conservative form
\begin{equation}
\partial_t \left(\gamma^{1/2}D\right) + \partial_i\left[\gamma^{1/2}\left(-\beta^i D + \alpha v^i D\right)\right] = 0,
\label{eq:mass2}
\end{equation}
\begin{equation}
 \begin{aligned}
\partial_t \left(\gamma^{1/2}S_j\right) & +  \partial_i\left[\gamma^{1/2}\left(-\beta^i S_j + \alpha {W^i}_j \right)\right] = \\ 
&  \gamma^{1/2}\left(\frac{1}{2} \alpha W^{ik} \partial_j \gamma_{ik} + S_i \partial_j \beta^i - U\partial_j\alpha\right),
\end{aligned}
\label{eq:momentum2}
\end{equation}
\begin{equation}
 \begin{aligned}
\partial_t \left(\gamma^{1/2}\tau\right) & +  \partial_i\left[\gamma^{1/2}\left(-\beta^i \tau + \alpha \left(S^i - v^i D\right)\right)\right] = \\
& \gamma^{1/2}\left(\frac{1}{2} W^{ik} \beta^j \partial_j \gamma_{ik} + {W^j}_i \partial_j \beta^i - S^j\partial_j\alpha\right),
\end{aligned}
\label{eq:energy2}
\end{equation}
where we repeated the equations solved in (\citealt{BHAC}) for
ideal-GRMHD. The purely spatial variant of the stress-energy tensor
$W^{ij}$ reads
\begin{equation}
 \begin{aligned}
  W^{ij} & \coloneqq {\gamma^i}_{\mu} {\gamma^j}_ {\nu} T^{\mu \nu} \\
& = \rho h \Gamma^2 v^i v^j - E^i E^j - B^i B^j + \left[p+ \frac{1}{2}\left(E^2 +B^2\right)\right]\gamma^{ij},
\end{aligned}
\label{eq:spatialstressenergy}
\end{equation}
with $h = h(\rho,p)$ the specific enthalpy of the fluid, $E^2:=E^i E_i$,
$B^2:=B^i B_i$ and $E^i$ and $B^i$ the three-vector spatial parts of the
electric and magnetic field in the Eulerian frame as defined in Eq.
(\ref{eq:Efieldeulerian}).

Equations (\ref{eq:mass2})--\eqref{eq:energy2} describe the evolution of
\textit{conserved quantities} as measured from an Eulerian reference
frame, namely the rest-mass density
\begin{equation}
D \coloneqq -\rho u^{\mu} n_{\mu} = \rho \Gamma,
\label{eq:density}
\end{equation} 
the covariant 3-momentum density
\begin{equation}
S_i \coloneqq {\gamma^{\mu}}_i n^{\alpha} T_{\alpha \mu} = \rho h \Gamma^2 v_i + \gamma^{1/2} \eta_{ijk} E^j B^k,
\label{eq:spatialmomentum}
\end{equation}
and the (rescaled) conserved energy density $\tau\coloneqq U-D$, where
\begin{equation}
U \coloneqq T^{\mu \nu} n_{\mu} n_{\nu} = \rho h \Gamma^2 - p +\frac{1}{2}\left(E^2+B^2\right).
\label{eq:energy1}
\end{equation}
The electric and magnetic fields are evolved through Maxwell's equations
described in Sec. \ref{sec:maxwelltheory}. In the absence of gravity,
when $\alpha =1$, $\beta^i = 0$, $\gamma^{1/2}=1$, and $\partial_t
\gamma= 0 $, these reduce to the special-relativistic conservation
laws. The non-relativistic (Newtonian) limit is obtained by letting $v^2
\ll 1$, $p \ll \rho$ and $E^2 \ll B^2 \ll \rho$, bearing in mind that
$c=1$.

Again, we emphasize that, unlike in ideal-GRMHD, the electric field in
equations (\ref{eq:momentum2}) and (\ref{eq:energy2}) cannot be
substituted as $E^i = -\gamma^{-1/2}\eta^{ijk}{B}_j v_k$ (with
$\eta_{ijk}$ the spatial Levi-Civita antisymmetric symbol). In the
resistive-GRMHD limit, $E^i$ has to be obtained from Amp\`{e}re's law
(see next Section), resulting in a larger system of equations and
therefore implying a more complex solution procedure.

%
%

\subsection{The Maxwell equations}
\label{sec:maxwelltheory}
The covariant Maxwell equations in tensorial form are
\begin{eqnarray}
&\nabla_{\nu}F^{\mu\nu} = \mathcal{J}^{\mu},\\
\label{eq:maxwell1}
&\nabla_{\nu} {}^*F^{\mu\nu} = 0,
\label{eq:maxwell2}
\end{eqnarray}
where $ \mathcal{J}^{\mu}$ is the electric 4-current. Applying the $3+1$
split, the tensors in the Maxwell equations (\ref{eq:maxwell1}) and
(\ref{eq:maxwell2}) can be decomposed in terms of the electromagnetic
fields as seen by an observer moving along the normal direction $n^{\nu}$
as
\begin{equation}
F^{\mu\nu} = n^{\mu}E^{\nu} - n^{\nu}E^{\mu} -(-g)^{-1/2} \eta^{\mu\nu\lambda\kappa}n_{\lambda} B_{\kappa},
\label{eq:maxwell12}
\end{equation}
\begin{equation}
{}^*F^{\mu\nu} = n^{\mu}B^{\nu} - n^{\nu}B^{\mu} +(-g)^{-1/2} \eta^{\mu\nu\lambda\kappa} n_{\lambda}E_{\kappa}.
\label{eq:maxwell22}
\end{equation}
with $\eta^{\mu\nu\lambda\kappa}$ the fully anti-symmetric symbol (see
e.g., \citealt{RezollaBook}) and electric and magnetic field four-vectors.
\begin{equation}
E^{\mu} := F^{\mu\nu}n_{\nu}, \qquad B^{\mu} := {}^*F^{\mu\nu}n_{\nu},
\label{eq:Efieldfour}
\end{equation}
and their three-vector spatial parts in the Eulerian frame 
\begin{equation}
E^{i} = F^{i\nu}n_{\nu} = \alpha F^{i0}, \qquad B^{i} = {}^*F^{i\nu}n_{\nu} = \alpha {}^*F^{i0}.
\label{eq:Efieldeulerian}
\end{equation}
Equation (\ref{eq:maxwell2}) can be written in component form, resulting
in Faraday's law,
\begin{equation}
\partial_t\left(\gamma^{1/2}B^j\right) + \partial_i\left[\gamma^{1/2}\left( \beta^jB^i - \beta^i B^j + \gamma^{-1/2} \eta^{ijk} \alpha E_k \right)\right] = 0.
\label{eq:faraday2}
\end{equation}
The temporal component of Eq. \eqref{eq:maxwell2} leads to the solenoidal
constraint
\begin{equation}
\gamma^{-1/2} \partial_i \left(\gamma^{1/2} B^i\right) = 0.
\label{eq:solenoidalconstraint}
\end{equation}
In addition to Faraday's law, Eq. \eqref{eq:maxwell1} can be written
in component form, resulting in Amp\`{e}re's law for the electric field
evolution in GRRMHD
\begin{equation}
\begin{aligned}
\partial_t\left(\gamma^{1/2}E^j\right) & + \partial_i\left[\gamma^{1/2}\left(\beta^j E^i - \beta^i E^j - \gamma^{-1/2} \eta^{ijk}\alpha B_k\right)\right] =  \\
& -\gamma^{1/2}\left(\alpha J^j - q\beta^j\right).
\end{aligned}
\label{eq:ampere2}
\end{equation}
The current density $\mathcal{J}^{\mu}$ is decomposed as
\begin{equation}
\mathcal{J}^{\mu} = n^{\mu}q + J^{\mu}
\label{eq:current1}
\end{equation}
where $J^{\mu} n_{\mu} = 0$, $q = -\mathcal{J}^{\mu}n_{\mu}$ is the
charge density, and $J^{\mu}$ the current density as measured by a
Eulerian observer moving with four-velocity $n^{\mu}$.  The temporal
component of Eq. \eqref{eq:maxwell1} then provides the charge
density $q$ in equation (\ref{eq:current1})
\begin{equation}
\gamma^{-1/2} \partial_i \left(\gamma^{1/2}E^i\right) = q.
\label{eq:charge}
\end{equation}
The spatial current density is obtained from the resistive Ohm's law in
$3+1$ split formulation (see e.g., \citealt{Palenzuela2};
\citealt{Bucciantini}).
\begin{equation}
J^i = q v^i + \frac{\Gamma}{\eta} \left[ E^i + \gamma^{-1/2} \eta^{ijk} v_j B_k - \left(v_k E^k\right)v^i\right],
\label{eq:current2}
\end{equation}
with the resistivity $\eta$ (not to be confused with the fully
anti-symmetric symbol $\eta^{\mu\nu\lambda\kappa}$), as the reciprocal of
the electrical conductivity, i.e., $\eta = 1/\sigma$. Substituting
equations (\ref{eq:charge}) and (\ref{eq:current2}) in (\ref{eq:ampere2})
we obtain the final form of Amp\`{e}re's law as
\begin{equation}
 \begin{aligned}
\partial_t\left(\gamma^{1/2}E^j\right) & + \partial_i\left[\gamma^{1/2}\left(\beta^j E^i - \beta^i E^j - \gamma^{-1/2} \eta^{ijk}\alpha B_k\right)\right] = \\
 & -\gamma^{1/2}\frac{\alpha \Gamma}{\eta} \left[ E^j + \gamma^{-1/2}\eta^{jik} v_i B_k - \left(v_k E^k\right)v^j\right] \\
& -\left(\alpha v^j - \beta^j\right) \partial_j\left(\gamma^{1/2}E^j\right).
\end{aligned}
\label{eq:ampere3}
\end{equation}
Note that the resistivity $\eta$ can depend both on space and time and
that this description of the current density is valid in any metric. Hall
or dynamo terms can be added by extending equation (\ref{eq:current2}) to
a generalized Ohm's law (e.g., \citealt{Bucciantini};
\citealt{Palenzuela}).

Finally, it is useful to introduce the fluid-frame (comoving) electric
and magnetic field (\citealt{Bucciantini}),
\begin{equation}
 e^\mu = \Gamma(E^i v_i)n^\mu + \Gamma(E^\mu + \gamma^{-1/2}\eta^{\mu\nu\lambda}v_\nu B_\lambda),
\end{equation}
\begin{equation}
 b^\mu = \Gamma(B^i v_i)n^\mu + \Gamma(B^\mu - \gamma^{-1/2}\eta^{\mu\nu\lambda}v_\nu E_\lambda),
\end{equation}
allowing to rewrite the electromagnetic part $T^{\mu\nu}_\mathrm{EM}$ of
Eq. (\ref{eq:stressenergy}) as (\citealt{Qian2016})
\begin{equation}
\begin{aligned}
 T^{\mu\nu}_\mathrm{EM} & = (b^2+e^2)\left(u^\mu u^\nu + \frac{1}{2}g^{\mu\nu}\right) - b^\mu b^\nu - e^\mu e^\nu \\
 & - u_\lambda e_\beta b_\kappa \left(u^\mu \gamma^{-1/2}\eta^{\nu\lambda\beta\kappa} + u^\nu \gamma^{-1/2}\eta^{\mu\lambda\beta\kappa}\right).
 \end{aligned}
\end{equation}
The comoving electric and magnetic field strength $e^2\coloneqq e^\mu
e_\mu$, $b^2\coloneqq b^\mu b_\mu$ are also employed in the definition of
useful dimensionless plasma quantities, e.g., the magnetization
$\sigma_\mathrm{mag}:=b^2/\rho$ and the gas-to-magnetic pressure ratio,
or plasma-$\beta_\mathrm{th}:=p_{\rm gas}/p_{\rm mag}=2p/b^2$.

\section{Numerical Implementation}
\label{sect:numericalgrrmhd}

In this Section we present the numerical approach to solve the set of
GRRMHD equations in \texttt{BHAC}. For small resistivity, the timescales
of the stiff and non-stiff parts of the system become very different and
the set of equations can be regarded as a hyperbolic system with
relaxation terms. These relaxation terms require special care to be
captured accurately without adopting an extremely small time-step. We
present and test the implementation of an IMEX RK method in
\texttt{BHAC}, where the stiff terms are treated with an implicit step
and the non-stiff parts with a standard explicit step. Our implementation
differs from previous works in the use of a new primitive-recovery method
(see Sec. \ref{sect:con2prim}) designed to obtain high accuracy and
reliability in regimes of low resistivity and low
plasma-$\beta_\mathrm{th}$. We test our algorithm against several
analytic and non-analytic benchmarks in Sec. \ref{sect:tests}. The
methods as presented in curved spacetime are straightforwardly applicable
in flat spacetime and the difficulties regarding stiff source terms are
completely analogous.

\subsection{The full system of equations in  {\tt BHAC} }
\label{sec:recap}
To adopt a conservative scheme we write the full system of equations
treated in \texttt{BHAC} in the form
\begin{align}
\partial_{t} (\gamma^{1/2} \, \boldsymbol{U}) + \partial_{i} (\gamma^{1/2} \, \boldsymbol{F}^{i}) = \gamma^{1/2} \, \boldsymbol{S} \,, \label{eq:conservationlaw}
\end{align}
where $\boldsymbol{U}$ represents conserved variables and
$\boldsymbol{F}^i$ are the fluxes,
\begin{align}
\boldsymbol{U} = 
\left[
\begin{array}{c}
D  \\
S_{j}  \\
\tau \\
B^{j} \\
E^{j}
\end{array}
\right] \,, \ \quad 
\boldsymbol{F}^{i} = 
\left[
\begin{array}{c}
\mathcal{V}^{i} D \\
\alpha W^{i}_{j} - \beta^{i} S_{j} \\
\alpha (S^{i}-v^{i} D) - \beta^{i} \tau \\
\beta^jB^i - \beta^i B^j + \gamma^{-1/2} \eta^{ijk} \alpha E_k \\
\beta^j E^i - \beta^i E^j - \gamma^{-1/2} \eta^{ijk}\alpha B_k
\end{array}
\right] \,, \label{eq:uandf}
\end{align}
with the transport velocity $\mathcal{V}^{i} :=\alpha v^{i} - \beta^{i}$. The sources read
\begin{align}
\boldsymbol{S} = 
\left[
\begin{array}{c}
0  \\
\frac{1}{2}\alpha W^{ik}\partial_{j}\gamma_{ik} + S_{i}\partial_{j}\beta^{i} - U\partial_{j}\alpha \\
\frac{1}{2} W^{ik} \beta^{j} \partial_{j} \gamma_{ik} + W_{i}^{j}\partial_{j}\beta^{i} - S^{j} \partial_{j} \alpha \\ 
0 \\
-\alpha J^j + \beta^{j}  q
\end{array}
\right] \,. \label{eq:ss}
\end{align}
The form of the GRRMHD equations as evolved in {\tt BHAC} allows for a
temporally and spatially dependent scalar resistivity $\eta(x^i,t)$. In
our implementation of GRRMHD the resistivity can depend on any dynamic or
static quantity, e.g., rest-mass density, current density or the position
explicitly (see \citealt{ripperda2019} for an application of non-uniform
current-dependent resistivity).

\subsection{Characteristic speed}
\label{sect:charspeed}
The characteristic velocities are required by the Riemann solver and the
Courant-Friedrichs-Lewy (CFL) condition that limits the time-step. Given the 3+1 structure
of the fluxes, we obtain characteristic waves of the form
\begin{align}
\lambda^i_\pm = \alpha \lambda^{\prime i}_\pm - \beta^i,
\end{align}
with $\lambda^{\prime i}_\pm$ the characteristic velocity in the $i$-th direction in the locally flat frame $\alpha\to 1,\, \beta^j\to 0$ (\citealt{anile1989}; \citealt{delzanna2007}). For simplicity we assume the characteristics to be in the limit of maximum diffusivity, i.e., the fastest waves locally travel with the speed of light, which after transforming to the Eulerian frame (\citealt{pons1998}; \citealt{white2016}) yields for each component (\citealt{delzanna2007}; \citealt{Bucciantini})
\begin{align}
\lambda^{\prime i}_\pm = \pm \sqrt{\gamma^{ii}}.
\end{align}
Note that a multidimensional Riemann solver for the SRRMHD equations was
recently presented by \cite{mignone2018,bodomignone} and
\cite{miranda2018}.

\subsection{Constraint equations}
\label{sec:constraints}

Our implementation of the GRRMHD equations enforces Eq.
(\ref{eq:solenoidalconstraint}) to roundoff-error by means of the
staggered CT scheme of \cite{balsara1999}, whose implementation in {\tt
  BHAC} has been presented in detail by \citet{Olivares2019}. The charge
density is obtained by numerically taking the divergence of the evolved
electric field as in Eq. (\ref{eq:charge}).

\subsection{Time stepping: IMEX method}
When the resistivity of the plasma is very small yet finite, the system
of Eqs. \eqref{eq:conservationlaw} becomes stiff. An explicit
integration, which is commonly used in ideal-GRMHD codes, then requires
time-steps that essentially scale with the resistivity, resulting in
prohibitive computational costs. In \cite{Komissarov}, a Strang-splitting
technique is applied in SRRMHD simulations such that the stiff resistive
terms can be explicitly computed for the electric field
evolution. However, the procedure relies on the assumption that magnetic
field and the fluid velocity field remains constant during the (faster)
evolution of the resistive electric field $E^i$. For small values of
$\eta$, the solution becomes inaccurate or requires extremely small
time-steps. An alternative solution is to split off the stiff part of the
system of Eqs. \eqref{eq:conservationlaw} and treat it with an implicit
step. With this approach no assumptions are necessary and in principle
all resistivity regimes can be treated without time-step restrictions
other than a standard CFL condition. This method was first proposed by
\cite{Palenzuela2} for SRRMHD and later extended by \cite{Bucciantini}
and \cite{Dionysopoulou} to GRRMHD. Several improvements of the IMEX
method for SRRMHD have been proposed recently by \cite{bodomignone}.

Here, we adopt the first-second order IMEX scheme as proposed by
\cite{Bucciantini}. In particular, we split the system
\eqref{eq:conservationlaw} into non-stiff
\begin{equation}
\partial_{t} (\gamma^{1/2} \, \boldsymbol{X}) = \boldsymbol{Q_X}\left(\gamma^{1/2}\boldsymbol{X},\gamma^{1/2}\boldsymbol{Y}\right),
\label{eq:conservationlaw_imex_nonstiff}
\end{equation}
and stiff equations
\begin{equation}
\partial_{t} (\gamma^{1/2} \, \boldsymbol{Y}) = \boldsymbol{Q_Y}\left(\gamma^{1/2} \boldsymbol{X},\gamma^{1/2} \boldsymbol{Y}\right) + \frac{1}{\eta} \boldsymbol{R_Y}\left(\gamma^{1/2} \boldsymbol{X},\gamma^{1/2} \boldsymbol{Y}\right), \label{eq:conservationlaw_imex_stiff}
\end{equation}
with the conserved quantities $\boldsymbol{U}$ split into two subsets
$\left\{\boldsymbol{X}, \boldsymbol{Y}\right\}$
\begin{align}
\boldsymbol{X} :=
\left[
\begin{array}{c}
D \\
S_i\\
\tau\\
B^i
\end{array}
\right] \,
, \ \qquad 
\boldsymbol{Y} := 
\left[
\begin{array}{c}
E^{i}
\end{array}
\right] \,, \label{eq:XandY}
\end{align}
containing the non-stiff and the stiff variables, respectively.

The time stepping involves a second-order time discretization for the
non-stiff variables in $\boldsymbol{X}$, which are evolved explicitly as
in \cite{BHAC}, and a first-order scheme for the stiff variables in
$\boldsymbol{Y}$, evolved implicitly. The overall solution step from time
level $n$ to $n+1$ is written
\begin{equation}
 \begin{aligned}
  & \tilde{\boldsymbol{X}}^{(1)} = \tilde{\boldsymbol{X}}^n + \frac{\Delta t}{2} \boldsymbol{Q_X}(\tilde{\boldsymbol{X}}^n,\tilde{\boldsymbol{Y}}^n), \\
  & \tilde{\boldsymbol{Y}}^{(1)} = \tilde{\boldsymbol{Y}}^n + \frac{\Delta t}{2} \boldsymbol{Q_Y}(\tilde{\boldsymbol{X}}^n,\tilde{\boldsymbol{Y}}^n) + \frac{\Delta t}{2\eta}\boldsymbol{R_Y}(\tilde{\boldsymbol{X}}^{(1)},\tilde{\boldsymbol{Y}}^{(1)}), \\
  & \tilde{\boldsymbol{X}}^{n+1} = \tilde{\boldsymbol{X}}^n + \Delta t \boldsymbol{Q_X}(\tilde{\boldsymbol{X}}^{(1)},\tilde{\boldsymbol{Y}}^{(1)}), \\
  & \tilde{\boldsymbol{Y}}^{n+1} = \tilde{\boldsymbol{Y}}^n + \Delta t \boldsymbol{Q_Y}(\tilde{\boldsymbol{X}}^{(1)},\tilde{\boldsymbol{Y}}^{(1)}) + \frac{\Delta t}{\eta}\boldsymbol{R_Y}(\tilde{\boldsymbol{X}}^{n+1},\tilde{\boldsymbol{Y}}^{n+1}), \\
 \end{aligned}
 \label{eq:imex12}
\end{equation}
where we have incorporated the $\gamma^{1/2}$ factors in
$\tilde{\boldsymbol{X}} := \gamma^{1/2}\boldsymbol{X}$,
$\tilde{\boldsymbol{Y}} := \gamma^{1/2}\boldsymbol{Y}$.

The implicit step represented by the $\boldsymbol{R_Y}$ terms can be
treated analytically due to the linearity (in the electric field) of the
resistive Ohm's law \eqref{eq:current2}. For simplicity, but without loss
of generality, consider the last electric field update step in the
algorithm above, from intermediate level $(1)$ to the next time level
$n+1$. Using Eq. \eqref{eq:current2}, this can be written explicitly
as
\begin{equation}
\begin{aligned}
\tilde{E}^{i,n+1} = \tilde{E}^{i,*} + \frac{\alpha\Delta t\Gamma}{\eta}& \left[ \tilde{E}^{i,n+1}+\gamma^{-1/2}\eta^{ijk}v^{n+1}_j\tilde{B}^{n+1}_k \right. \\
& \left. - \left(\tilde{E}^{k,n+1}v^{n+1}_k\right)v^{i,n+1}\right],
\label{eq:implicitupdateE}
\end{aligned}
\end{equation}
where $\tilde{E}^i:=\gamma^{1/2}E^i$ and
$\tilde{B}^i:=\gamma^{1/2}B^i$. Here, the explicitly updated electric
field is $\tilde{E}^{i,*} = \tilde{E}^{i,n} + \Delta t
\boldsymbol{Q_Y}(\tilde{\boldsymbol{X}}^{(1)},
\tilde{\boldsymbol{E}}^{(1)})$. Equation \eqref{eq:implicitupdateE} only
involves local operations (no spatial derivatives needed), hence its
inversion is straightforward if the terms on the right-hand-side are
known, and leads to an explicit expression for the new electric field,
\begin{equation}
\begin{aligned}
 \tilde{E}^{i,n+1} =  \frac{\tilde{E}^{i,*}}{1+\sigma_\mathrm{H}\Gamma^{n+1}} & - \frac{\Gamma^{n+1}}{1+\sigma_\mathrm{H}\Gamma^{n+1}} \Bigg[\gamma^{-1/2}\eta^{ijk}v^{n+1}_j \tilde{B}^{n+1}_k \\
 & \left. - \Gamma^{n+1}\frac{\tilde{E}^{k,*}v_k^{n+1}}{\Gamma^{n+1}+\sigma_\mathrm{H}}v^{i,n+1}\right],
\label{eq:explicitupdateE1}
\end{aligned}
\end{equation}
where $\sigma_\mathrm{H}:=\alpha\Delta t/\eta$. In order to avoid
singularities in the ideal-MHD limit $\eta \rightarrow 0$, the equation
above can be recast as
\begin{equation}
\begin{aligned}
 \tilde{E}^{i,n+1} =  \frac{\eta\tilde{E}^{i,*}}{\eta+\sigma_\mathrm{L}\Gamma^{n+1}} & - \frac{\sigma_\mathrm{L}\Gamma^{n+1}}{\eta+\sigma_\mathrm{L}\Gamma^{n+1}} \Bigg[\gamma^{-1/2}\eta^{ijk}v^{n+1}_j \tilde{B}^{n+1}_k \\
 & \left. - \eta\Gamma^{n+1}\frac{\tilde{E}^{k,*}v_k^{n+1}}{\eta\Gamma^{n+1}+\sigma_\mathrm{L}}v^{i,n+1}\right],
\label{eq:explicitupdateE2}
\end{aligned}
\end{equation}
where $\sigma_\mathrm{L}:=\alpha\Delta t = \eta \sigma_\mathrm{H}$. Note
that \cite{Bucciantini} have a typo in their equations (33), (35), and
(36) for the formulation of the implicit and explicit updates. The
analytic inversion is applied in the same way at each substep of the
time-stepping algorithm, hence making Eqs. \eqref{eq:explicitupdateE1}
and \eqref{eq:explicitupdateE2} completely general by adjusting
$\sigma_\mathrm{H}$ and $\sigma_\mathrm{L}$ with the coefficients from a
Butcher tableau corresponding to the current substep
(\citealt{pareschirusso2005}; \citealt{Palenzuela2}). Therefore, the
simple first-second order IMEX algorithm \eqref{eq:imex12} can be
extended to arbitrary high-order accuracy while keeping the update
equations for $E^i$ unchanged. However, contrary to higher-order schemes,
the first-second algorithm \eqref{eq:imex12} naturally includes the
ideal-MHD limit, $\eta=0$, without suffering from numerical singularities
(\citealt{Bucciantini}).

Note that the electric field update in Eqs. \eqref{eq:explicitupdateE1}
and \eqref{eq:explicitupdateE2} involves the three-velocity $v^i$ to be
known at the same time level of $E^i$. However, $v^i$ is a primitive
quantity which depends nonlinearly on $E^i$. This dependence makes the
update equations intrinsically implicit, and implies that the electric
field update in the IMEX algorithm \eqref{eq:imex12} must be carried out
concurrently to a conserved-to-primitive variable inversion. The
inversion recovery strategy is of key importance for GR(R)MHD simulations
and high sensitivity to the physical parameters makes it a particularly
challenging part of the algorithm.

\subsection{Transformation of conserved to primitive variables}
\label{sect:con2prim}

Throughout the solution process of the GRRMHD equations
\eqref{eq:conservationlaw}, a transformation of the \textit{conserved
  variables} $D$, $S_i$, $\tau$ into the \textit{primitive variables}
$\rho$, $v_i$, and $p$ is necessary. This is a local operation that
requires to solve the system of nonlinear Eqs. \eqref{eq:density},
\eqref{eq:spatialmomentum}, and \eqref{eq:energy1}. The solution of such
a system cannot be written in closed form, requiring a root-finding
algorithm that constitutes one of the most expensive and sensitive parts
of the whole solution procedure of relativistic MHD codes\footnote{Note
  that $B^i$ is both a conserved and a primitive variable, hence an
  inversion step is not needed for the magnetic field.}.

For most of the operations during the GRMHD evolution (i.e., as long as
the electric field does not depend on primitive variables), we carry out
the conserved-to-primitive inversion by solving the system of equations
\begin{equation}
\begin{aligned}
& D := \rho\Gamma, \\
& S_i := \xi v_i + \gamma^{1/2}\eta_{ijk} E^j B^k, \\
& \tau := \xi - p - D +\frac{1}{2}\left(E^2 + B^2\right),
\end{aligned}
\label{eq:c2phydro}
\end{equation}
where $\xi:=\rho h \Gamma^2$. Provided that $E^i$ and $B^i$ are known,
such a system can usually be reduced to one single equation (in $\xi$,
$p$, or other scalar variables, see e.g., \citealt{noble2006} or
\citealt{siegel2018}) and solved with a one-dimensional (1D)
Newton-Raphson (NR) iteration (where 1D refers to the single scalar
equation that has to be solved and not to a spatial dimension). This is
the standard approach in \texttt{BHAC} for the solution of the
ideal-GRMHD equations (\citealt{vanderholst_2008};
\citealt{keppens_2012}; \citealt{BHAC}).

However, the IMEX scheme presented in the previous Section for the GRRMHD
equations involves an implicit update of $E^i$ where both the new
electric field and the new three-velocity are unknown. In this case, the
system of nonlinear Eqs. \eqref{eq:c2phydro} above cannot be inverted, as
$E^i$ is not known a priori but rather an additional variable determined
by Eq. \eqref{eq:explicitupdateE1} or
\eqref{eq:explicitupdateE2}. Therefore, during the implicit step the
conserved to primitive transformation must be carried out concurrently to
the implicit electric field update. The system of equations
\eqref{eq:c2phydro} is thus augmented with equation
\eqref{eq:explicitupdateE1} or \eqref{eq:explicitupdateE2} for the
electric field, forming again a closed set in the variables $\rho$,
$v_i$, $\xi$, and $E^i$. The new system requires a robust and accurate
nonlinear solution method, typically an iterative algorithm. This
combined update-transformation step is a crucial operation, which heavily
influences the overall performance and accuracy of the IMEX algorithm. If
the iteration fails to converge, the electric field cannot be updated and
the GRRMHD solution becomes inaccurate. The high failure rate in this
step is an issue reported in several relativistic resistive MHD
implementations, particularly in low plasma-$\beta_\mathrm{th} \lesssim
0.5$ regimes (\citealt{Palenzuela2}; \citealt{Dionysopoulou};
\citealt{delzanna2016}; \citealt{Qian2016}). A robust inversion method
that is reliable in particularly demanding physical regimes (e.g.,
low-$\beta_\mathrm{th}$, high-$\sigma_\mathrm{mag}$) is essential to
model accretion flows onto black holes.

Here we present a set of strategies for the inversion-update step. Based
on the performance of each strategy, we design a robust approach that
yields a minimal amount of failures, allowing for a wide range of
simulation parameters that are unattainable with currently available
methods.

\subsubsection{``1D'' fixed-point strategy}
\label{sect:c2p1D}
The most commonly used approach in GRRMHD consists of reducing the system
of nonlinear equations to one (\citealt{Palenzuela2};
\citealt{Dionysopoulou}), or two (\citealt{Bucciantini};
\citealt{Qian2016}) scalar equation(s). A usual choice is to solve the
energy equation
\begin{equation}
 \xi = p + D + \tau - \frac{1}{2}\left(E^2(v_i)+B^2\right)
 \label{eq:residualxi}
\end{equation}
for the scalar variable $\xi$, hence the ``1D'' fixed-point notation, or alternatively ``2D'' fixed-point for two scalar variables (see e.g., \citealt{noble2006} and \citealt{Qian2016} for an iteration on $\Gamma$ and $W := (\rho + p\hat{\gamma}/(\hat{\gamma}-1))\Gamma^2$). The pressure $p(\rho,\xi)$ is determined
by eliminating the dependence on $\rho$ by substitution with
$\rho=D/\Gamma$, and reduced to a function of $\xi$ only via the relation
\begin{equation}
 \Gamma = \sqrt{1 + \frac{{S'}^2}{\xi^2-{S'}^2}},
\end{equation}
which follows from Eq. \eqref{eq:c2phydro} above. Here,
${S'}^2:=S'^iS'_i$, with $S'_i := S_i - \gamma^{1/2}\eta_{ijk}E^j B^k$. The
dependence of $E^i$ on $v_i$ [Eqs. \eqref{eq:explicitupdateE1} or
  \eqref{eq:explicitupdateE2}], however, is nonlinear and cannot be
recast into an explicit relation $E^i(\xi)$. As a consequence, the usual
solution approach consists of a hybrid NR iteration where the dependence
of $E^i$ on $\xi$ is not taken into account, and the electric field is
obtained with a fixed-point iteration. Starting from an initial guess for
$\xi$ and $E^i$, each nonlinear iteration is composed of the following
steps:
\begin{enumerate}
\item Compute the velocity as
  \begin{equation}
  v_i = \frac{S_i-\gamma^{1/2}\eta_{ijk}E^{j}B^k}{\xi^{(m)}}.
  \end{equation}
\item Compute $\Gamma$ and $p$, and the electric field from equation
  \eqref{eq:explicitupdateE1} or \eqref{eq:explicitupdateE2}.
 \item Compute the residual,
   \begin{equation}
     f(\xi) = \xi - p - D - \tau +\frac{1}{2}(E^2+B^2),
   \end{equation}
   and its derivative neglecting the dependence of $E^i$ on $\xi$,
   \begin{equation}
     \frac{df}{d\xi} = 1 - \frac{dp}{d\xi}.
   \end{equation}
 \item Update the value of $\xi$ at the $m-$th iteration with a NR step,
   \begin{equation}
     \xi^{(m+1)} = \xi^{(m)} - f(\xi^{(m)}) \left(\frac{df(\xi^{(m)})}{d\xi}\right)^{-1}.
   \end{equation}
 \item Track the absolute change in the iteration variables,
   $|\xi^{(m+1)}-\xi^{(m)}|$ and $|E^{i,(m+1)}-E^{i,(m)}|$. The iteration
   is stopped if this difference falls below a prescribed tolerance,
   which we normally take to be $10^{-14}$.
\end{enumerate}

Step 3 above is where a crucial assumption is introduced. Computing the
electric field with a fixed-point strategy of this type and neglecting
the dependence $E^i(\xi)$ is equivalent to assuming that the electric
field only varies slightly between successive Newton steps. This is not
always true, especially when the system is very stiff. The stiffness of
the nonlinear equations can originate from a parameter choice (e.g., for
low values of the resistivity $\eta$), or from the physical regime
described by the conserved quantities (e.g., large electromagnetic energy
density compared to the rest-mass density or pressure resulting in low
$\beta_\mathrm{th}$ and high $\sigma_\mathrm{mag}$). In such cases, the
electric field becomes dynamically important, and its variation with
respect to other quantities cannot be neglected.

Most implementations of IMEX schemes employing the 1D (or 2D) fixed-point
scheme above report numerical issues related to combinations of
low-$\eta$, high-$\sigma_\mathrm{mag}$, and low-$\beta_\mathrm{th}$
regimes (\citealt{Palenzuela2}; \citealt{Qian2016}). Failures in the
inversion-update step can sometimes be mitigated by reducing the time
step (\citealt{Palenzuela2}), which effectively reduces the stiffness in
the nonlinear system of equations to invert. However, this is an
undesirable constraint especially for production runs of accretion flows,
where large regions of low-$\beta_\mathrm{th}$ (i.e.\ the ambient
surrounding the accretion disk) or high-$\sigma_\mathrm{mag}$ (i.e.\ the
jet) can rapidly determine a degradation of computational performance.

\subsubsection{``3D/4D'' fully-consistent strategies}
A more robust approach to the inversion-update problem is to eliminate
any assumption on the importance of the dynamics of $E^i$, and treat the
whole system of nonlinear equations simultaneously and consistently. The
system of equations \eqref{eq:c2phydro} for the fluid variables, together
with the relation $\Gamma=(1-v^2)^{-1/2}$, involves in principle 6
independent unknowns; the augmented system including equation
\eqref{eq:explicitupdateE1} or \eqref{eq:explicitupdateE2} for the
electric field, increases the number of total unknowns to 9. It is
essential to reduce the problem to a smaller set of equations, in order
to improve the robustness of the iterative solution procedure. A larger
number of unknowns implies a higher computational cost and involves a
larger solution space, thus decreasing the likelihood of convergence. In
our analysis, we find that the problem can be reduced to a minimal system
of three or four scalar equations, depending on the quantities chosen as
iteration variables.

In general, the system of equations is described by a set of nonlinear
residuals $\textbf{f}(\textbf{x})$ in the unknowns $\textbf{x}$, which
contains either three or four components (hence 3D/4D). Starting from an
initial guess, we adopt an iterative strategy that progressively
decreases the residuals until $\textbf{f}(\textbf{x})\simeq
\mathbf{0}$. In \texttt{BHAC}, the iteration is typically carried out
with a multi-dimensional NR scheme (using a hardcoded analytic Jacobian
$\textbf{H}(\textbf{x})=\partial\textbf{f}(\textbf{x})/\partial\textbf{x}$). For
robustness and flexibility, we have the option of selecting a
Jacobian-free Newton-Krylov (NK) scheme that does not require the full
Jacobian but only directional derivatives, thus allowing for new
strategies to be easily implemented (see e.g., \citealt{kelley} for a
reference implementation of NK schemes).

In our analysis of the inversion-update equations, we find that the
iterative scheme yields the lowest failure rates when applied to the
following reduced systems:
\begin{itemize}
 \item \textit{3D system on $u_i$:} since the new electric field is an
   explicit function of $v_i$ [Eq. \eqref{eq:explicitupdateE1} or
     \eqref{eq:explicitupdateE2}], we find that it is possible to reduce
   the iteration to the three components of the fluid velocity, recast as
   the normalized 3-momentum $u_i=\Gamma v_i$, with
   $\Gamma=\sqrt{1+\gamma^{ij}u_i u_j}$. The iteration variables are then
   $\textbf{x}=\textbf{u}=(u_1,u_2,u_3)$ and the residuals take the form
 \begin{equation}
  \textbf{f}(\textbf{u}) = u_i - \frac{S_i - \gamma^{1/2}\eta_{ijk}E^j B^k}{Dh},
  \label{eq:residual3du}
 \end{equation}
where $E^j$ is determined by Eq. \eqref{eq:explicitupdateE1} or
\eqref{eq:explicitupdateE2} as a function of $\textbf{u}$, and
$h(\rho,p)$ is computed as a function of $\textbf{u}$ only, by
substituting $\rho=D/\Gamma$ and by recasting $p=p(\textbf{u})$. The
latter approach may not be possible for an arbitrary or tabulated equation of state (EoS); 
however, this operation is straightforward for the EOS choices
available in \texttt{BHAC} and typically employed for simulations of
accretion flows onto black holes. For instance, the closure equation for
perfect fluids with polytropic index $\hat{\gamma}$ can be written as
$p=\rho(\hat{\gamma}-1)\epsilon$, where
\begin{equation}
  \epsilon = \Gamma \frac{\tau'}{D} - z \frac{\sqrt{{S'}^2}}{D} + \frac{z^2}{1+\Gamma}
\end{equation}
with ${\tau'}:=\tau - (E^2+B^2)/2$ and $z^2 := \Gamma^2-1$. With all
quantities written as explicit functions of the iteration variables
$\textbf{u}$, the Newton algorithm can be
applied to minimize the residuals \eqref{eq:residual3du}. We note that a similar
3D approach has been presented by \cite{Bucciantini} and \cite{bodomignone}.
 
\item \textit{4D system on ($\xi$,$u_i$):} as an alternative to the 3D
  system above, we can choose to retain $\xi$ as an additional unknown
  related to the energy of the system. The iteration variables in this
  case are $\textbf{x}=(\xi,\textbf{u})$ and the residuals read
 \begin{equation}
 \textbf{f}(\xi,\textbf{u}) = \left[
 \begin{array}{c}
 \xi - p - D - \tau + \left(E^2 + B^2\right)/2 \\
 u_i - \Gamma \left(S_i - \gamma^{1/2}\eta_{ijk}E^j B^k\right)/\xi
 \end{array}
 \right] .
 \label{eq:residual4dxiu}
 \end{equation}
 Here, the electric field is still computed from $\textbf{u}$, and the
 pressure is determined from $\xi$ and $\Gamma$ as in the 1D approach of
 Sec. \ref{sect:c2p1D}. This strategy involves a larger system of
 equations to handle compared to the 3D case above, but requires less
 operations at each nonlinear iteration.
 
\item \textit{4D system on ($z$,$E^i$):} as a final alternative we recast
  the system of equations to a formulation involving $z$ and the electric
  field $\textbf{E}$ as iteration variables, $\textbf{x} =
  (z,\textbf{E})$. The residuals for this case are written
\begin{equation}
 \textbf{f}(z,\textbf{E}) = \left[
 \begin{array}{c}
 z - \sqrt{{S'}^2}/(Dh) \\
 E^i - f^i_E(v_i)
 \end{array}
 \right],
 \label{eq:residual4dze}
\end{equation}
where $f^i_E(v_i)$ are the right-hand sides of equation
\eqref{eq:explicitupdateE1} or \eqref{eq:explicitupdateE2}. The velocity
is explicitly computed from the iteration variables as
$v_i=S'_i/(Dh\Gamma)$, and the specific enthalpy is retrieved from the
pressure $p(z,\textbf{E})$ similarly to the 3D strategy above.
 
\end{itemize}

By considering several possible strategies for the inversion-update step,
we are allowed to explore a wide range of important properties such as
computational cost and convergence rate in the parameter space of
interest. As a general approach, in the schemes above we prefer to rely
on variables that are not constrained between specific limiting values
due to physical consistency, e.g., we make use of $u_i \in
(-\infty,\infty)$ rather than $v_i \in (-1,1)$, or $z \in (-\infty,\infty)$, which is preferable to
  $\Gamma \in [1,\infty)$ (\citealt{galeazzi}). In this way, the
    iterative scheme is less likely to fail due to the variables assuming
    out-of-range values. Additionally, the availability of several
    schemes allows for designing a robust backup strategy: in the case of
    failure of a primary inversion scheme, a second one can be employed
    that relies on different variables. The choice of iteration variables
    affects the convergence properties of each scheme, and we show in
    Sec. \ref{sect:unittests} how a backup strategy can be designed
    such that the smallest number of failures is achieved. As initial
    guess for the iterative schemes, we typically employ the value of the
    unknown quantities at the previous time-step (we find that different
    initial guesses produce little variations in the overall
    performance). When convergence is reached, we check the final values
    for all primitive quantities for physical consistency.

\subsubsection{Entropy inversion}
\label{sect:entropy}
In highly magnetized regions of accretion flows (e.g., the relativistic
jet in accretion flow simulations), the evolution equation for the
conserved energy $\tau$ can sometimes become too numerically inaccurate,
resulting in unphysical solutions to the conserved to primitive inversion
problem. In such situations, \texttt{BHAC} relies on an additional backup
strategy for the conserved to primitive inversion, based on the entropy
$\kappa$. For most of the code operations that involve a standard
inversion step (i.e.\ when $E^i$ is known at the current time-step), this
entropy ``switch'' consists of finding the root of a single nonlinear
equation, typically in the unknown $\Gamma$. The energy $\tau$ is
discarded in the process, and replaced with a value consistent with the
newly recovered primitives. For details on this standard procedure we
refer the reader to the corresponding Section in \cite{BHAC}.

When the entropy-switch is needed during the more complicated
inversion-update step in the implicit part of our IMEX scheme, the system
cannot be reduced to one single equation. We approach the problem by
applying the same 3D/4D strategies presented above, with slight
modifications. For the 3D scheme in $u_i$ and the 4D scheme in $(z,E^i)$,
we replace any closure relation with an ideal-gas law for the enthalpy
(\citealt{RezollaBook})
\begin{equation}
 h = 1+\frac{\hat{\gamma}}{\hat{\gamma}-1}\frac{p}{\rho},
\end{equation}
augmented with the polytropic (isentropic) EOS
$p=\kappa\rho^{\hat{\gamma}}$. The pressure can still be computed
explicitly from the iteration variables, therefore leaving the scheme
essentially unchanged. For the 4D scheme in $(\xi,\textbf{u})$, we
replace the equation for $\xi$ such that the residuals read
\begin{equation}
 \textbf{f}(\xi,\textbf{u}) = \left[
 \begin{array}{c}
 \xi - \rho h \Gamma^2 \\
 u_i - \Gamma \left(S_i - \gamma^{1/2}\eta_{ijk}E^j B^k\right)/\xi
 \end{array}
 \right],
 \label{eq:residual4dxiuentropy}
\end{equation}
where the enthalpy is still given by the polytropic EOS as above.

If the entropy-switch strategy is activated, it is applied by default
whenever the primary inversion fails. Particularly for accretion flow
simulations, the entropy-switch is also applied upon successful primary
inversion in regions where a low $\beta_\mathrm{th}<10^{-2}$ is
detected. Switching between different systems of nonlinear equations,
which is required for the entropy-based inversion, is easily handled in
\texttt{BHAC} with the NK subroutines, which do not require the full
Jacobian for each different inversion strategy. In case of failure of the
entropy-based inversion, the last-resort solutions include the
replacement of the primitive variables in the faulty cells with averages
from nearby converged zones. Alternatively, floor values in pressure,
velocity, and rest-mass density can be set if convergence in the
inversion step is not reached. Note that the electric field is not
floored to an arbitrary value, but rather recalculated with equation
\eqref{eq:explicitupdateE1} or \eqref{eq:explicitupdateE2} by using the
floor value of $v_i$.


\section{Numerical tests}
\label{sect:tests}

In this Section we present a number of validation tests for our
implementation of the GRRMHD algorithm. We include comparisons on the
reliability of the inversion-update strategies presented in Sec.
\ref{sect:con2prim}, based on which we design a robust strategy that is
employed as our default choice in \texttt{BHAC}. We also analyse the
performance of the entropy-based inversion scheme with dedicated
tests. We present one- and two-dimensional case studies and compare the
results to reference high-resolution runs and analytic
solutions. Finally, we show an example application of our algorithm in
the astrophysically relevant case of a torus accreting onto a black hole.

\subsection{Exploration of parameter space for primitive-variable recovery}
\label{sect:unittests}

In order to test the inversion-update strategies presented in Sec.
\ref{sect:con2prim}, we explore the performance of each algorithm in a
large parameter space. We choose to study vastly different GRMHD regimes
with $\sigma_\mathrm{mag} \in [10^{-2},10^2]$ and $\beta_\mathrm{th}\in
[10^{-10}, 10^5]$, by selecting appropriate sets of primitive
variables. By picking primitive quantities $\Gamma-1\in [10^{-2}, 10^3]$
and $\eta \in [10^{-14}, 10^6]$, we can construct complete sets of
corresponding conserved variables. These are then run through the
inversion-update algorithm, whose performance can be evaluated by
comparing the retrieved primitives with the manufactured initial
sets. Our calculations are performed in flat spacetime, hence without
considering the effect of spacetime curvature onto the inversion-update
process.

The 3D-$u_i$, 4D-$(\xi,u_i)$, and 4D-$(z,E^i)$ inversion schemes are
applied with and without the entropy-switch strategy discussed in
Sec. \ref{sect:entropy}. These results are compared to the performance of
the 1D-$\xi$ scheme with fixed-point calculation of $E^i$. Our comparison
includes the results of a ``backup'' strategy designed to yield the
maximum rate of successful inversions. This is constructed by applying
one of the three multi-D strategies (typically the 3D-$u_i$) and then
switching to another scheme upon failure of the inversion process. As a
final backup measure, the entropy-switch strategy in $u_i$ is called in
case the three primary strategies fail in retrieving a valid set of
primitives (for these tests, we do not apply the $\beta_\mathrm{th}$
threshold described in Sec.  \ref{sect:entropy}). This backup combined
strategy shows dramatic improvements over the standard 1D-$\xi$ scheme,
and is the default choice in \texttt{BHAC} for production runs. In the
unit tests below, the order of the strategies used in the ``backup''
approach is 3D-$u_i$, 4D-$(\xi,u_i)$, 4D-$(z,E^i)$, and finally the
entropy-switch. In all cases, we use a NR approach with
hardcoded Jacobian (we find no significant differences in the convergence
rates when applying a NK scheme instead).

As a first test, we explore the $(\eta,\sigma_\mathrm{mag})$ parameter
space. Considering a wide range of values for the resistivity allows for
investigating the importance of the dynamics of $E^i$ compared to the
ideal-GRMHD limit for both high and low magnetization. The corresponding
sets of primitives are constructed by choosing $B^2=1$, $\Gamma=2$, and
$\beta_\mathrm{th}=0.1$ as fixed parameters, as applicable for a
relativistic magnetized plasma. For the electric field update, we choose
an electric field strength $(E^*)^2=0.1$ (that is normally obtained from
an explicit update) and a time-step $\Delta t=0.01$. The polytropic index
is fixed to $\hat{\gamma}=2$.

\begin{figure*}
\centering
\includegraphics[width=2\columnwidth,trim={30mm 10mm 10mm 0mm},clip]{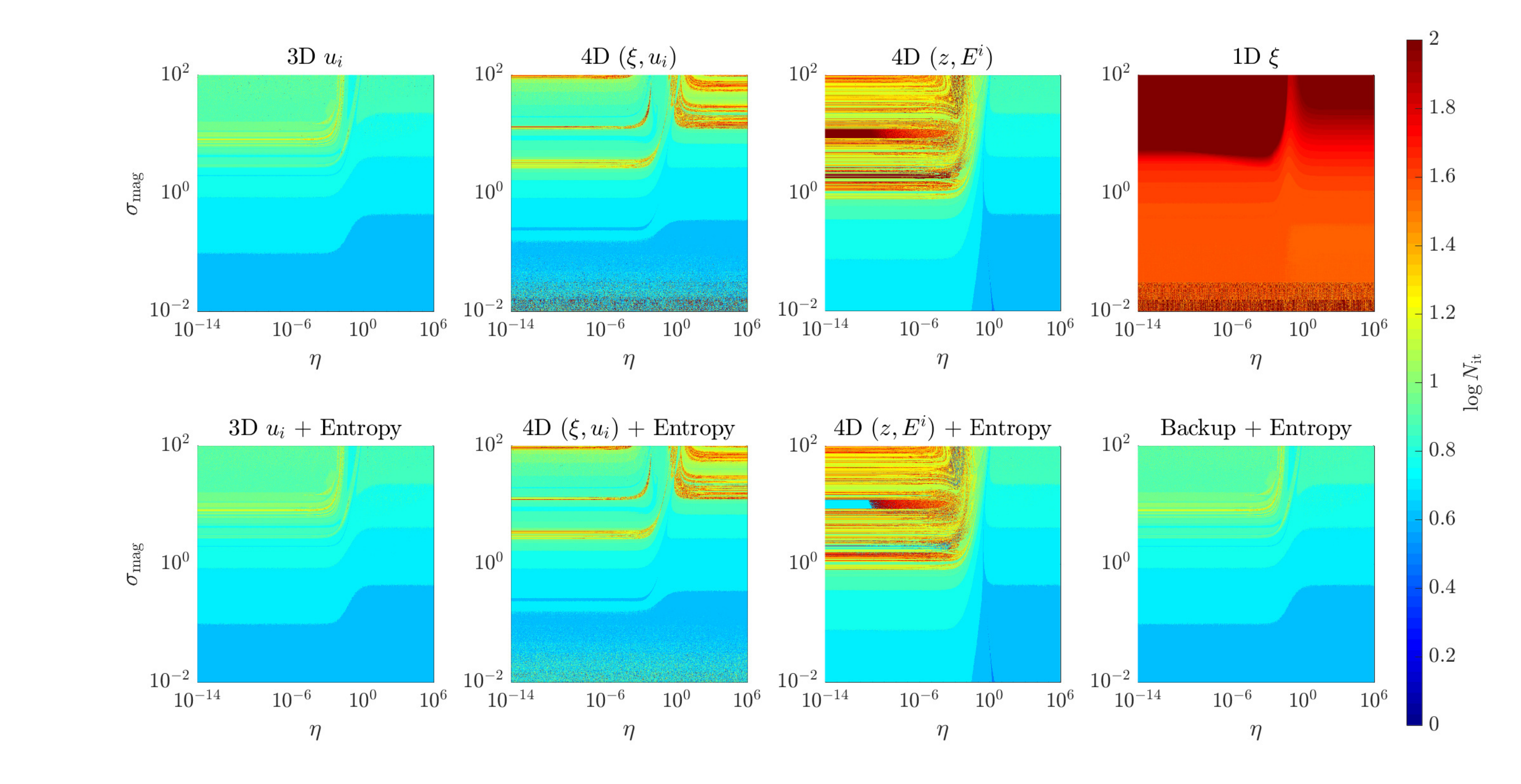}
\caption{Convergence plots for the inversion-update strategies applied to the
  $(\eta,\sigma_\mathrm{mag})$ parameter space, in terms of number of
  iterations needed (limited to 100). The manufactured sets of primitive
  variables are constructed by choosing $B^2=1$, $\Gamma=2$,
  $\beta_\mathrm{th}=0.1$, $(E^*)^2=0.1$, and $\Delta t=0.01$. Each panel 
  represents for $10^{6}$ conserved-to-primitive inversions. The
  inversion schemes are applied without (top row) and with (bottom row)
  entropy-switch as a backup strategy. The new combined ``backup'' scheme
  (bottom right) shows zero total failures, dramatically surpassing the
  performance of the 1D scheme (top right).}
\label{fig:c2psigmaeta}
\end{figure*}

Figure \ref{fig:c2psigmaeta} shows the results of $10^{6}$ conserved-to-primitive 
inversions expressed in terms of the
number of iterations needed to reach a solution of the nonlinear system
with an absolute accuracy of $10^{-14}$ in the computed primitives. The
maximum amount of iterations allowed is 100, after which the algorithm is
stopped and the inversion is considered as having failed (denoted by dark
red dots in the plots). Note that, in production runs, additional checks
are applied on the iteration error when the maximum iteration number 
(we typically allow for 100 iterations) is reached. In this
case, if the final error is only slightly larger than the prescribed
tolerance, the solution can still be considered valid via a larger,
user-defined acceptance tolerance. The two 4D strategies (middle columns)
show rather complementary regions of failure, with the $(\xi,u_i)$ scheme
being more reliable for high-$\sigma_\mathrm{mag}$ zones and the
$(z,E^i)$ scheme converging more easily for low-$\sigma_\mathrm{mag}$
zones. The 3D scheme in $u_i$ (leftmost column) shows the highest rate of
successful inversions, with no specific regions of failed recovery. The
entropy-switch applied to the three strategies as backup options (bottom
row) shows slightly larger success rate, but does not change the
convergence regions qualitatively. All strategies show superior
performance compared to the standard 1D strategy in $\xi$ with
fixed-point calculation of $E^i$ (top-right panel), both in terms of
convergence rate and number of required iterations. The combined
``backup'' strategy (our default choice for calculations in
\texttt{BHAC}) therefore provides a dramatic improvement over the often
applied fixed-point strategies, with zero failures in the explored
parameter space, compared to a $\sim13\%$ failure rate for the 1D scheme
in $\xi$, although the set of equations are admittedly slightly different
if the entropy-switch is applied.

\begin{figure*}
\centering \includegraphics[width=2\columnwidth,trim={30mm 10mm 10mm
    0mm},clip]{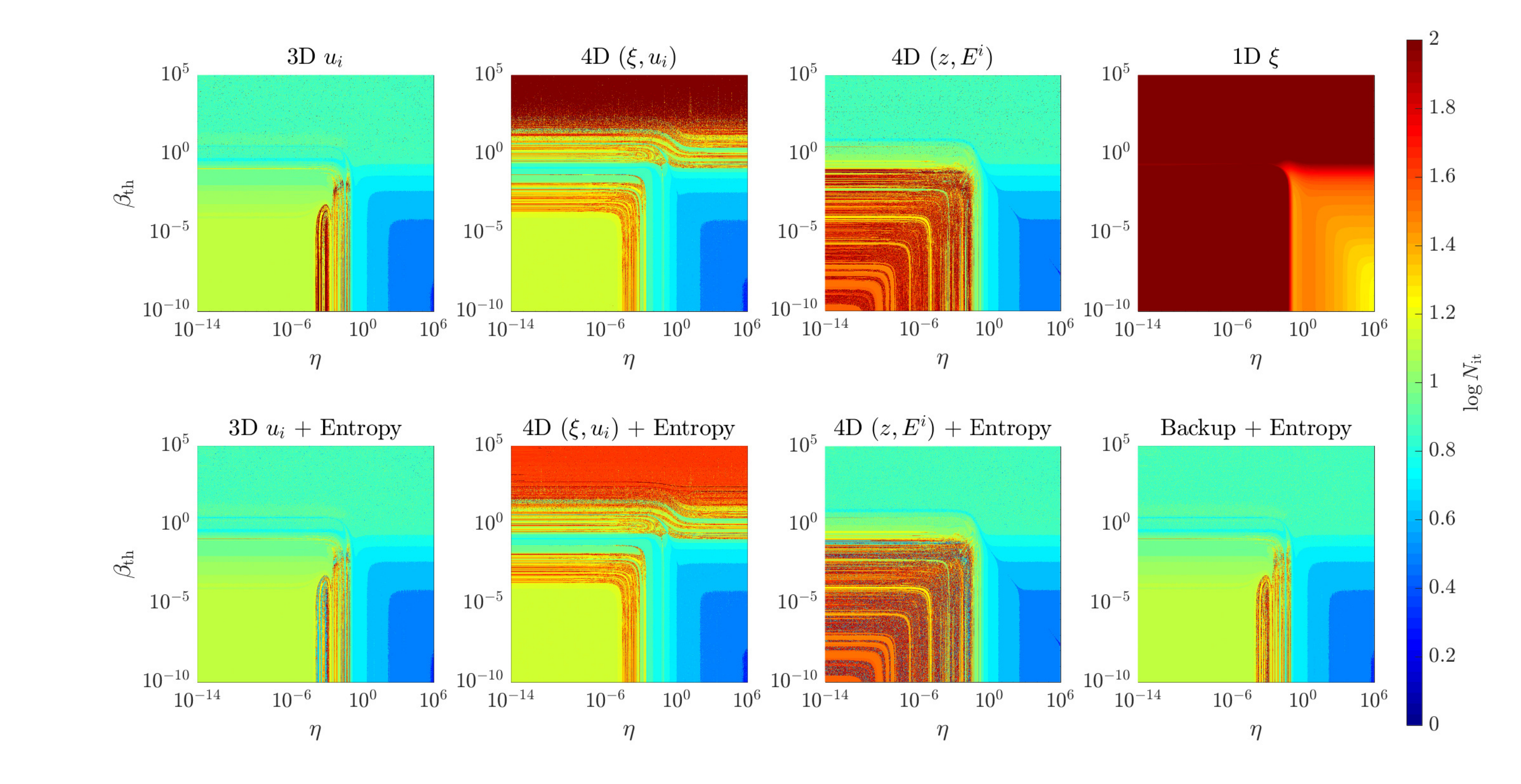}\caption{Convergence plots
  for the inversion-update strategies applied to the $(\eta,\beta_\mathrm{th})$
  parameter space, in terms of number of iterations needed (limited to
  100). The manufactured sets of primitive variables are constructed by
  choosing $B^2=1$, $\Gamma=2$, $\sigma_\mathrm{mag}=10$, $(E^*)^2=0.1$,
  and $\Delta t=0.01$. Each panel represents for $10^{6}$ conserved-to-primitive 
  inversions.The inversion schemes are applied without (top
  row) and with (bottom row) entropy-switch as a backup strategy. The new
  combined ``backup'' scheme (bottom right) shows a $\sim0.005\%$ failure
  rate, a major improvement over the $\sim76\%$ failure rate of the 1D
  scheme (top right).}
\label{fig:c2pbetaeta}
\end{figure*}

As a second test, we explore the $(\eta,\beta_\mathrm{th})$ parameter
space. Considering the variation of $\beta_\mathrm{th}$ relates the
resistive dynamics of $E^i$ (defined by $\eta$) with the case of
magnetically dominated (i.e., low-$\beta_\mathrm{th}$) or thermally
dominated energy (i.e., high-$\beta_\mathrm{th}$) plasma. The primitive sets 
are constructed upon fixing $B^2=1$, $\Gamma=2$, 
$\sigma_\mathrm{mag}=10$, $(E^*)^2=0.1$, and $\Delta
t=0.01$. The results are shown in Fig. \ref{fig:c2pbetaeta}, which
illustrates how the performance of the three new schemes is similar to
the previous case, with almost complementary convergence zones for the 4D
schemes, and seemingly scattered failures for the 3D scheme in $u_i$. For
this case, the entropy-switch greatly increases the success rate of the
inversion procedure. Overall, the new backup-strategy combined with
entropy-switch (bottom-right panel) yields a $\sim0.005\%$ failure rate,
a major improvement over the standard 1D scheme in $\xi$ (top-right
panel). The latter shows a large region of no convergence, with an
overall $\sim76\%$ failure rate. Failures in the fixed-point strategy
seem to be mostly driven by simultaneous conditions of low resistivity
and low-$\beta_\mathrm{th}$, which could preclude modelling large zones
of accretion flows which are magnetically-dominated and nearly-ideal
($\eta \rightarrow 0$).

\begin{figure*}
\centering
\includegraphics[width=2\columnwidth,trim={30mm 10mm 10mm 0mm},clip]{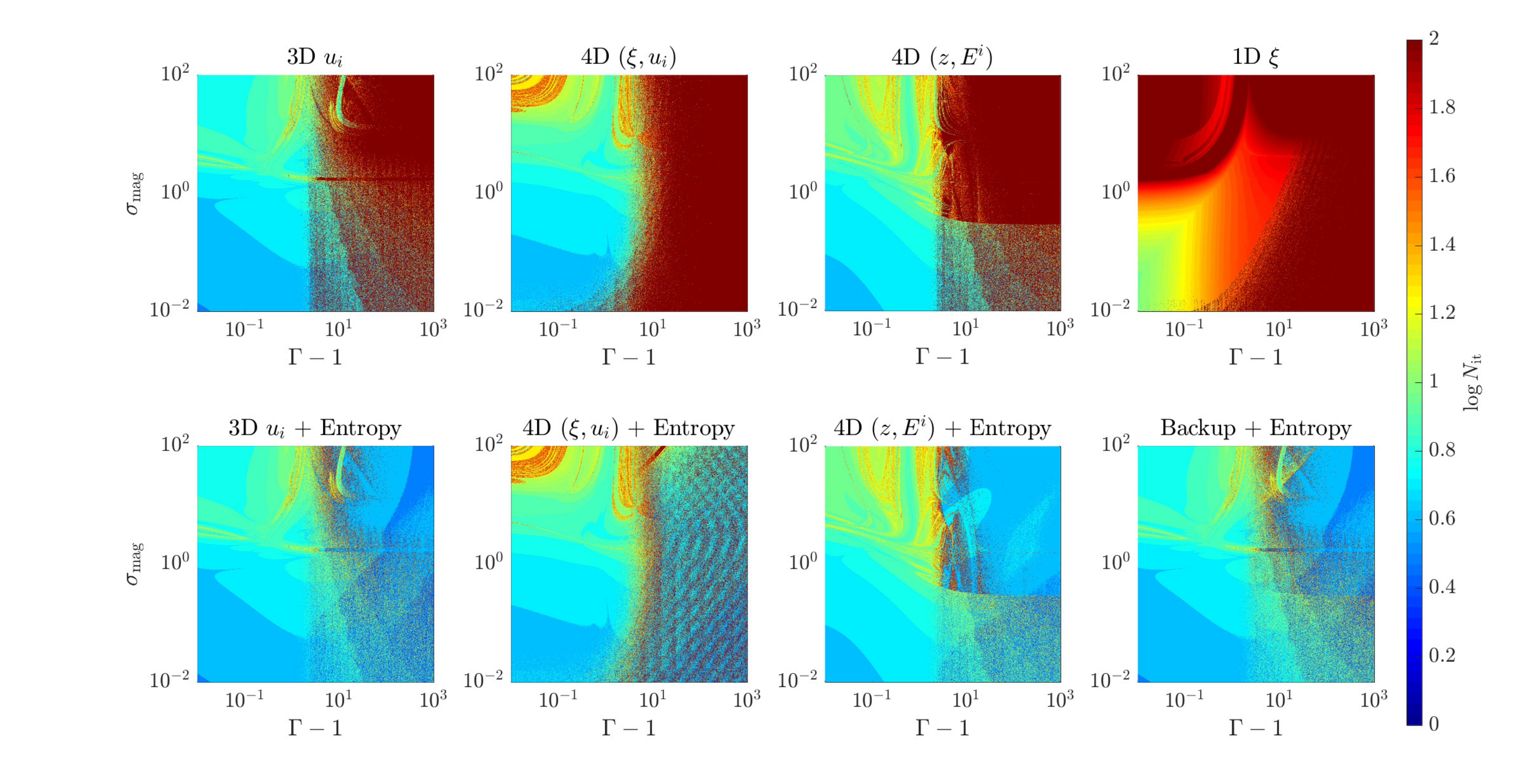}
\caption{Convergence plots for the
  inversion-update strategies applied to the
  $(\Gamma,\sigma_\mathrm{mag})$ parameter space, in terms of number of
  iterations needed (limited to 100). The manufactured sets of primitive
  variables are constructed by choosing $B^2=1$, $\eta=0.1$,
  $\beta_\mathrm{th}=0.1$, $(E^*)^2=0.1$, and $\Delta t=0.01$. Each panel 
  represents for $10^{6}$ conserved-to-primitive inversions. The
  inversion schemes are applied without (top row) and with (bottom row)
  entropy-switch as a backup strategy. The new combined ``backup'' scheme
  (bottom right) shows a $\sim0.007\%$ failure rate, with a major
  improvement over the $\sim48\%$ failure rate of the 1D scheme (top
  right).}
\label{fig:c2psigmalfac}
\end{figure*}

As a final test, we consider the $(\Gamma,\sigma_\mathrm{mag})$ parameter
space. High-$\sigma_\mathrm{mag}$, high-$\Gamma$ regions are common in
accretion flow simulations, e.g., in the highly magnetized jet emerging
from compact objects, where the fluid can be accelerated to high Lorentz
factors. Here we manufacture the primitive sets by fixing $B^2=1$,
$\eta=0.1$, $\beta_\mathrm{th}=0.1$, $(E^*)^2=0.1$, and $\Delta
t=0.01$. The results in Fig. \ref{fig:c2psigmalfac} show a generally
large region of failure at high-$\Gamma$ for all strategies. The
entropy-switch significantly improves the convergence rate for all the
strategies. Compared to the standard 1D strategy in $\xi$ (showing a
$\sim 48\%$ failure rate), our combined backup strategy shows failures
only in a restricted $\sim 0.007\%$ of the considered parameter space.

In summary: our tests clearly indicate that the new inversion-update
strategies presented in Sec. \ref{sect:con2prim} are necessary to
properly handle a wide range of regimes typically encountered in GR(R)MHD
simulations of accretion flows and of compact-binary
mergers. We observe dramatic improvements over the standard approach of a
1D inversion scheme on a scalar with fixed-point calculation of $E^i$, with
recorded failures only in an extremely limited number of cases (e.g., 49
failures over a total of $10^6$ points considered in the
$(\eta,\beta_\mathrm{th})$ space, or zero failures in the considered
$(\eta,\sigma_\mathrm{mag})$ space). However, one should be careful when
considering the entropy-switch as a reliable backup strategy: using the
entropy results in a different physical system, where a lower temperature
is assumed as applicable in the isentropic limit
$d(p/\rho^{\hat{\gamma}})/dt=0$. In all cases, we detect a final error on
the computed primitives of the same order of the iteration error (hence
below $10^{-14}$ in case of convergence).

\subsection{Shock-tube tests}
\label{sec:shocktube}
As a second test actually solving the set of GRRMHD equations, we have
considered a one-dimensional shock-tube in flat spacetime. Such tests are
very restrictive for code validation and show strong nonlinear behavior
and steep discontinuities. The ability of the code to handle a range of
resistivities for such a problem is essential for astrophysical
applications where shocks are ubiquitous. We use the shock-tube setup as
proposed by \cite{briowu} to test the code performance, and compare
to the results in the ideal-GRMHD limit from \cite{BHAC}. For nonzero resistivity we compare the efficiency and performance of the different inversion methods, including the benchmark Strang-split scheme of \cite{Komissarov}. Considering the lack of exact solutions for shock-tubes with nonzero resistivity, we postpone testing convergence properties to Sections \ref{sec:sheet}, \ref{sec:vortex}, and \ref{sec:bondi}.

The initial conditions are given by:
\begin{align}
&(\rho, p, v^x, v^y, v^z, B^x, B^y, B^z) = \\
&\left\{
\begin{array}{llllllllr}
(1.0,& 1.0,& 0.0,& 0.0,& 0.0,& 0.5,& \phantom{-}1.0,& 0.0) \qquad  &~~~~~x<0\nonumber \\
(0.125,& 0.1,& 0.0,& 0.0,& 0.0,& 0.5,& -1.0,& 0.0) \qquad  &~~~~~x>0\\
\end{array}
\right.
\label{eq:shocktubebhac}
\end{align}
with an adiabatic index of $\hat{\gamma} = 2$. These settings result in
$\beta_\mathrm{th} = 1.6$ for $x < 0$ and $\beta_\mathrm{th} = 0.16$ for
$x>0$. We use a uniform grid with 1024 points spanning $x \in [-1/2,
  1/2]$. We adopt a second-order TVD limiter (\citealt{koren1993}) for
spatial reconstruction with CFL number of 0.4.

All tests have been performed with both the Strang-split scheme of
\cite{Komissarov} and with all inversion-update methods for the IMEX
scheme. Note that for these tests we do not activate the
entropy fix, nor do we replace faulty cells or apply the floor
models since for the multi-D inversion strategies no failures are encountered.

\begin{figure*} 
\begin{center}
\subfloat{\includegraphics[width=0.5\textwidth]{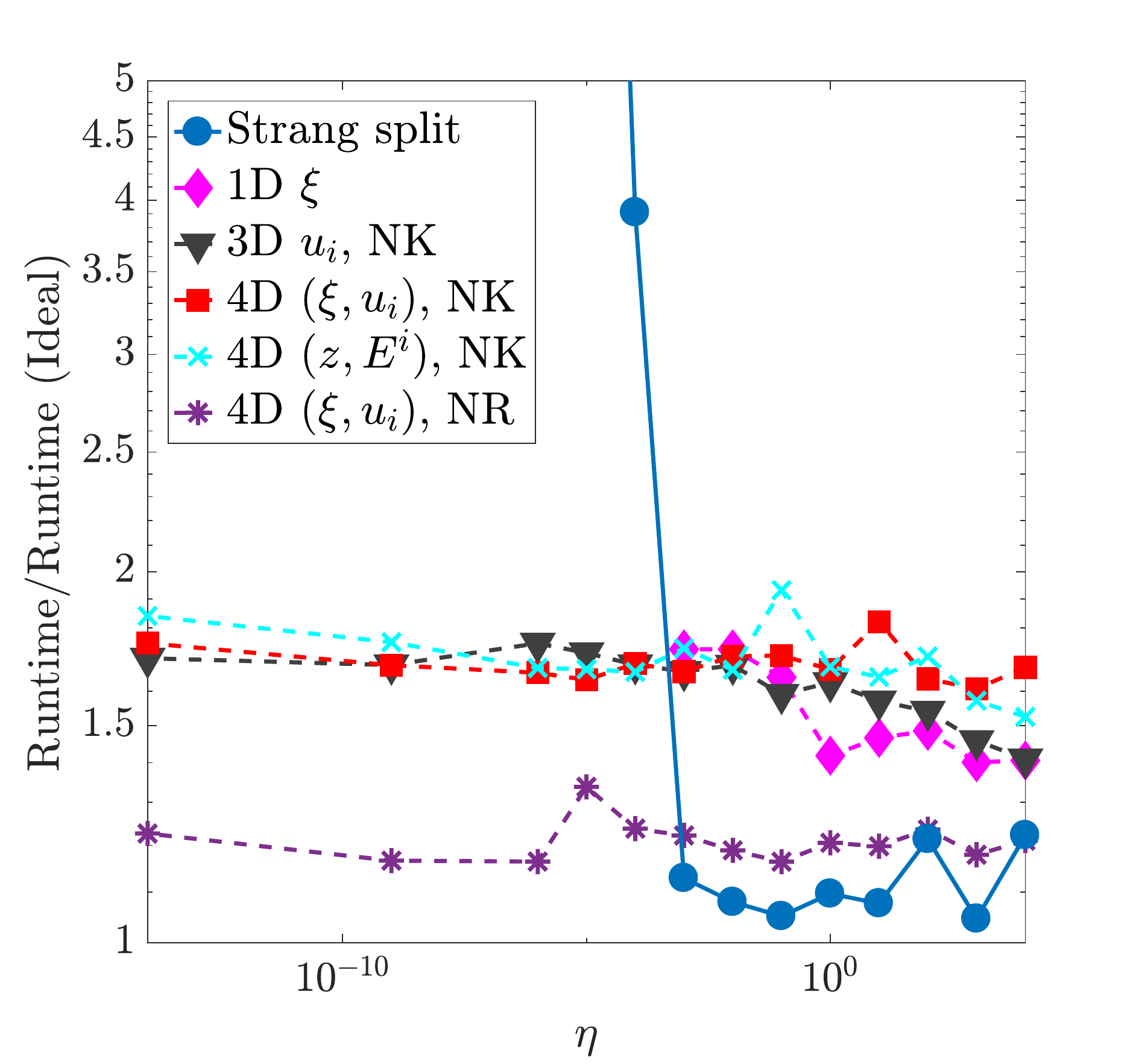}}
\subfloat{\includegraphics[width=0.5\textwidth]{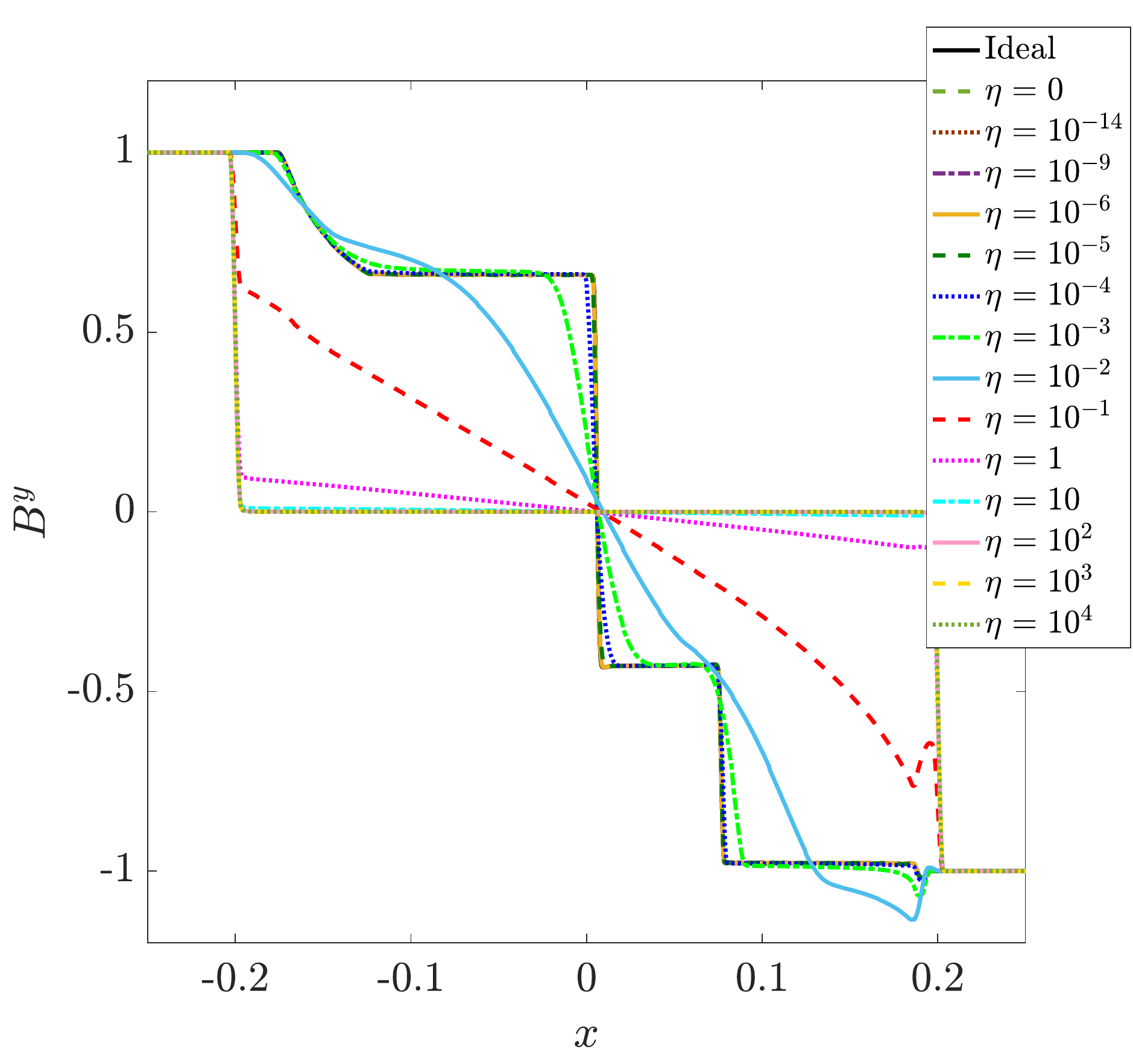}}
\caption{Shock-Tube test as in \protect\cite{briowu}. Left panel: runtime
  normalized by the runtime of ideal-GRMHD in {\tt BHAC} for all
  resistivities and a selected number of inversion schemes. Right panel:
  $B^y$ component of the magnetic field for a range of resistivities
  $\eta \in [0, 10^4]$. Also the ideal-GRMHD ($\eta=0$) result of
  \protect\cite{BHAC} is shown. The results for the different methods
  cannot be distinguished visually. Cases with $\eta \leq 10^{-6}$ have not been reproduced
  with the Strang-split method of \protect\cite{Komissarov} due to the
  extremely small time-step needed. For $\eta \leq 10^{-6}$ there is no
  visual difference between ideal-GRMHD and GRRMHD solutions.}
\label{fig:shocktube}
\end{center}
\end{figure*}

In the right panel of Fig. \ref{fig:shocktube}, the results for the
$B^y$-component of the magnetic field are shown for all resistivities
$\eta \in [0, 10^4]$ considered, at $t = 0.2$. The results obtained with
the IMEX and Strang schemes cannot be distinguished visually for $\eta
\geq 10^{-5}$ cases. For $\eta \geq 10$ the results correspond to the
zero conductivity case and for $\eta \leq 10^{-5}$ no visual differences
are observed between resistive and ideal-MHD.

In the left panel we compare the runtime for the different
primitive-recovery methods, normalized to the runtime of ideal-GRMHD. The
Strang split method of \cite{Komissarov} performs best for high
resistivity $\eta \geq10^{-3}$, since no additional iterations on the
electric field are included in the conserved to primitive
transformation. However, for $\eta < 10^{-3}$ the proportionality of the
time-step to the resistivity rapidly decreases the performance to
prohibitively long runtimes.

For the IMEX scheme with a 1D $\xi$ inversion method we observe a similar
trend, and without reducing the CFL condition no convergence is reached
for $\eta < 10^{-3}$. For example, the 1D $\xi$ inversion scheme needs a
CFL condition of 0.06 for $\eta = 10^{-4}$, a CFL condition of 0.006 for
$\eta = 10^{-5}$, and 0.0006 for $\eta = 10^{-6}$ (these results are not
shown in Fig. \ref{fig:shocktube}, where we keep the CFL
fixed). \cite{Palenzuela2} reached a similar conclusion, stating that the
1D primitive variable recovery for more demanding Riemann problems (such
as this shock-tube case) lacks robustness for ratios of
$\beta_\mathrm{th} \lesssim 0.4$. Note that the shock-tube as tested in
\cite{dumbser2009}, \cite{Bucciantini}, and \cite{Qian2016} is less
restrictive for the inversion scheme, due to the smaller discontinuity in
plasma-$\beta_\mathrm{th} \in [0.45, 0.4]$ between the left and right
state in the initial conditions. Our tests confirm that this less
demanding setup can be correctly modeled for any $\eta$ without reducing
the time-step with all IMEX inversion schemes (including the 1D $\xi$
method) considered in this work.

The 3D and 4D inversion schemes presented in Sec. \ref{sect:con2prim}
perform similarly and always produce the correct solution, with a runtime
that is comparable to the 1D method. The runtime always remains within a
factor $\sim 2$ of the runtime of the ideal-GRMHD solver in {\tt
  BHAC}. If using a hardcoded Jacobian for the primitive recovery, such
that a NR iterative method can be applied, these schemes become even
faster and always produce a correct solution for the shock-tube within
$\sim1.5$ times the ideal-GRMHD runtime, for any value of the
resistivity.

\subsection{Self-similar current sheet}
\label{sec:sheet}
The third test case is the evolution of a thin current sheet first
considered by \cite{Komissarov}. Once the layer has expanded over several
times its initial width, a self-similar evolution ensues.  The analytic
solution at time $t$ is described by
\begin{equation}
B^y(x,t) = {\rm erf} \left({\frac{x}{2\sqrt{\eta t}}}\right),
\end{equation}
for the magnetic field, while the electric field evolves as
\begin{equation}
E^z(x,t) = \sqrt{\frac{\eta}{\pi t}} \exp\left(-\frac{x^2}{4\eta t}\right),
\end{equation}
and we set $t=1$ as initial condition, to start with a resolved state in
the self-similar phase. Rest-mass density and pressure are homogeneous and
set to $\rho=1$ and $p=5000$, while all remaining GRRMHD variables are
set to zero. The dynamics takes place in the $x$-direction which is
resolved between $x\in[-1.5,1.5]$ by $256$ grid-points. Here we fix the
resistivity to $\eta=0.01$. The test is reproduced with all 3D and 4D
inversion methods. Note that for these tests we do not activate the
entropy fix, nor do we replace faulty cells or apply the floor models. In
all cases we apply an HLL reconstruction scheme with a Koren-type limiter
(\citealt{koren1993}) and we keep a CFL ratio of 0.5.

\begin{figure} 
\centering
\includegraphics[width=1\columnwidth]{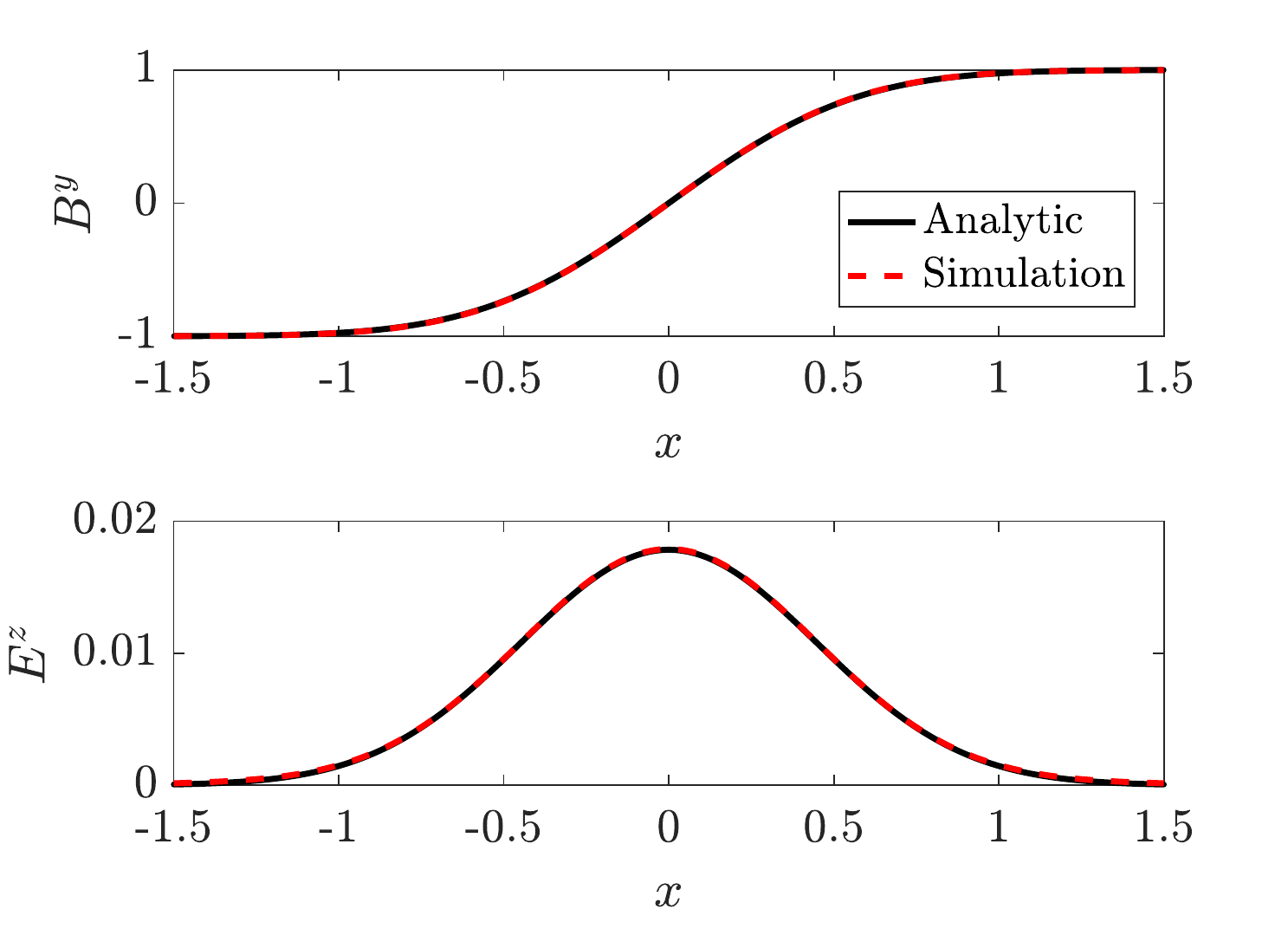}
\caption{Self-similar current sheet solution at $t=10$ as in
  \protect\cite{Komissarov} on 256 grid-points.}
\label{fig:currentsheet}
\end{figure}

\begin{figure} 
\centering
\includegraphics[width=1\columnwidth]{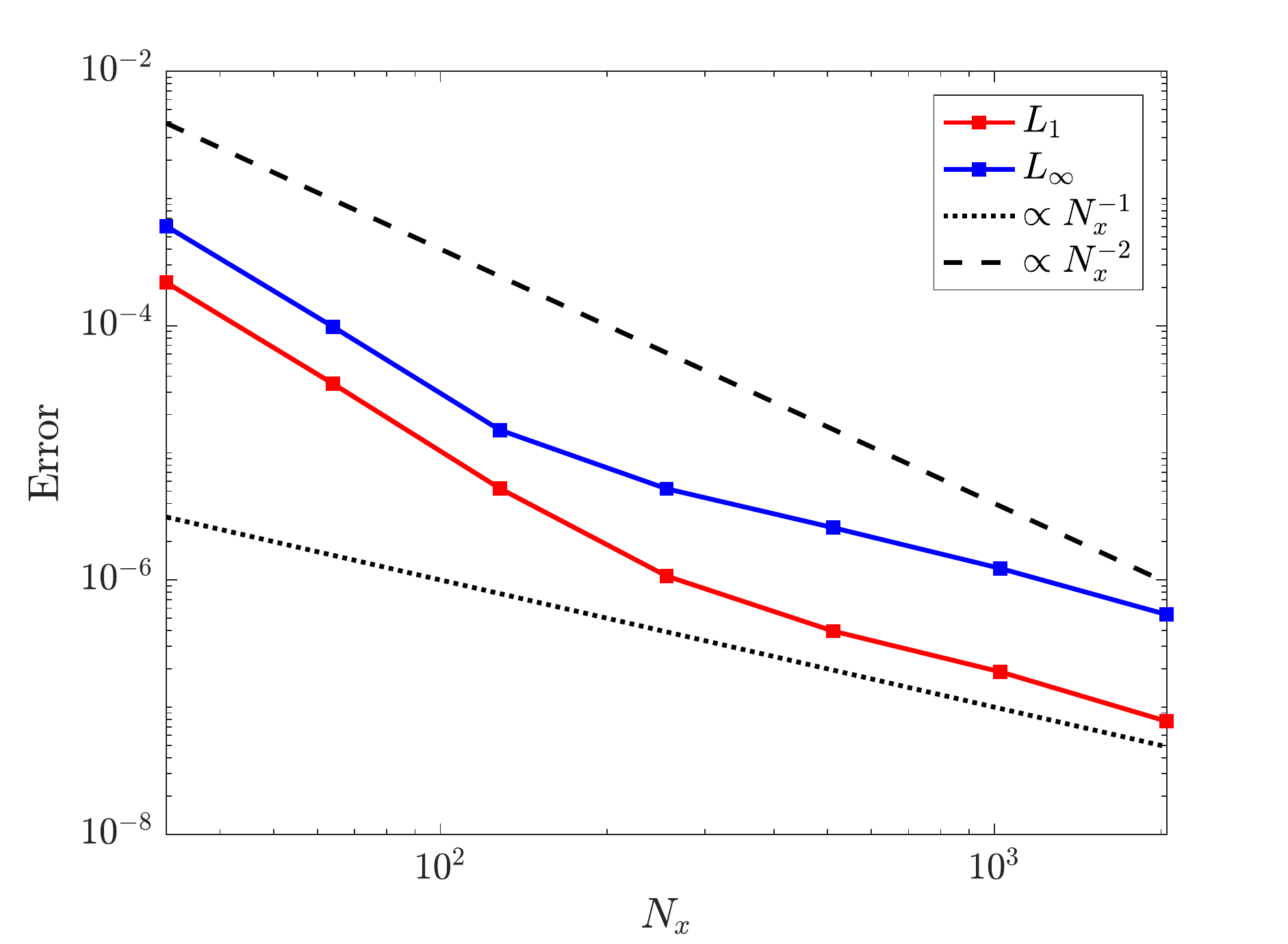}
\caption{Convergence study for the current sheet evolution at increasing
  number of grid-points $N_x$. The $L_1$ and $L_\infty$ norms of the
  difference between a high-resolution run and the numerical results
  indicate first-order convergence for high-resolution runs (where
  temporal discretization errors dominate over spatial errors), as
  expected from the properties of the first-second IMEX scheme by
  \protect\cite{Bucciantini}.}
\label{fig:currentsheetconv}
\end{figure}

The analytic solution for $B^y$ and $E^z$ is shown in Fig.
\ref{fig:currentsheet} at time $t=10$ (black line). The numerical results
(red line) cannot be distinguished visually. In order to assess the
accuracy of the evolution, we study the order of convergence of the
numerical solution. We measure the $L_1$ and $L_\infty$ norm of the error
in the numerical solution by progressively increasing the number of
grid-points and comparing to a high-resolution run with 8192
grid-points. We choose not to compare the numerical results with the
analytic solution above, which is only valid in the limit of infinite
pressure (as pointed out by \citealt{Bucciantini}). The error trend thus
obtained is reported in Fig. \ref{fig:currentsheetconv}. We observe that,
for low-resolution runs, the accuracy of the scheme is above second order
(as expected by the use of a Koren limiter for a smooth solution), a sign
that spatial errors dominate over temporal inaccuracies. For high
resolutions, where spatial errors become progressively less important,
the scheme tends to first-order accuracy, as is expected from the
application of the first-second order IMEX scheme from
\cite{Bucciantini}.

Finally, in order to test the implementation of the fluxes in $3+1$ split formulation, 
we run the setup under different gauges. These are summarized in table 
\ref{tab:funkycoord}. The results including gauge effects are indistinguishable from
Fig. \ref{fig:currentsheet} once the coordinate-transformations have been
accounted for.

\begin{table}
\centering
\caption{Coordinates for the gauge effect tests.}\label{tab:funkycoord}
\begin{tabular}{ccrrrr}
\hline
Case & $\alpha$ & $\beta^{i}$ & $\gamma_{11}$ & $\gamma_{22}$ & $\gamma_{33}$ \\
\hline
A & 1 & (0,0,0) & 1 & 1 &1
\\
B & 2 & (0,0,0) & 1 & 1 & 1
\\
C & 1 & (0.4,0,0) & 1 & 1 & 1
\\

D & 1 & (0,0,0) & 4 & 1 & 1
\\

E & 1 & (0,0,0) & 1 & 4 & 1
\\

F & 2 & (0.4,0,0) & 4 & 9 & 1
\\
\hline
\end{tabular}
\end{table}

\subsection{Charged vortex}
\label{sec:vortex}

\cite{bodomignone} has recently proposed the first exact two-dimensional
equilibrium solution of the SRRMHD equations, which describes a rotating
flow with a uniform rest-mass density in a vertical magnetic field and a
radial electric field. Adopting a set of cylindrical coordinates
$(r,\phi,z)$, the solution is given by
\begin{equation}
\begin{aligned}
& E_r = \frac{q_0}{2}\frac{r}{r^2+1}, \\
& B_z = \frac{\sqrt{(r^2+1)^2-q_0^2/4}}{r^2+1}, \\
& v_{\phi} = -\frac{q_0}{2}\frac{r}{\sqrt{(r^2+1)^2-q_0^2/4}},\\
& p = -\frac{\rho (\hat{\gamma}-1)}{\hat{\gamma}}+\left(p_0 + \frac{\rho(\hat{\gamma}-1)}{\hat{\gamma}}\right)\left(\frac{4r^2+4-q_0^2}{(r^2+1)(4-q_0^2)}\right)^{\frac{\hat{\gamma}}{2(\hat{\gamma}-1)}},
\end{aligned}
\end{equation}
with radial coordinate $r:=x^2+y^2$ and on the axis $r=0$, we choose the
charge density $q_0 = 0.7$, pressure $p_0 = 0.1$ and uniform rest-mass
density $\rho = 1$ in the whole domain in accordance with
\cite{bodomignone}.  The adiabatic index is set as $\hat{\gamma} = 4/3$
and the resistivity as $\eta = 10^{-3}$. Note that \cite{bodomignone}
evolve the charge density with a separate evolution equation and set it
initially as $q = q_0 / (r^2+1)^2$, whereas in {\tt BHAC} it is obtained
as the divergence of the evolved electric field [cf.,
  Eq. \eqref{eq:charge}].

The simulation is carried out on a two-dimensional Cartesian grid with
$x,y \in [-10,10]$ with a uniform resolution of $[N_x\times N_y]$ and
$(N_x=N_y=32,64,128,256,512)$ cells until time $t=5$. We apply continuous
extrapolation of all quantities at the boundaries. We use the 3D--$u_i$
primitive-recovery method from Sec. \ref{sect:con2prim} (similar to the
inversion method used by \citealt{bodomignone}) and we do not activate
the entropy fix, nor do we replace faulty cells or apply the floor
models since no inversion failures are encountered.

\begin{figure*} 
\begin{center}
\subfloat{\includegraphics[width=0.5\textwidth]{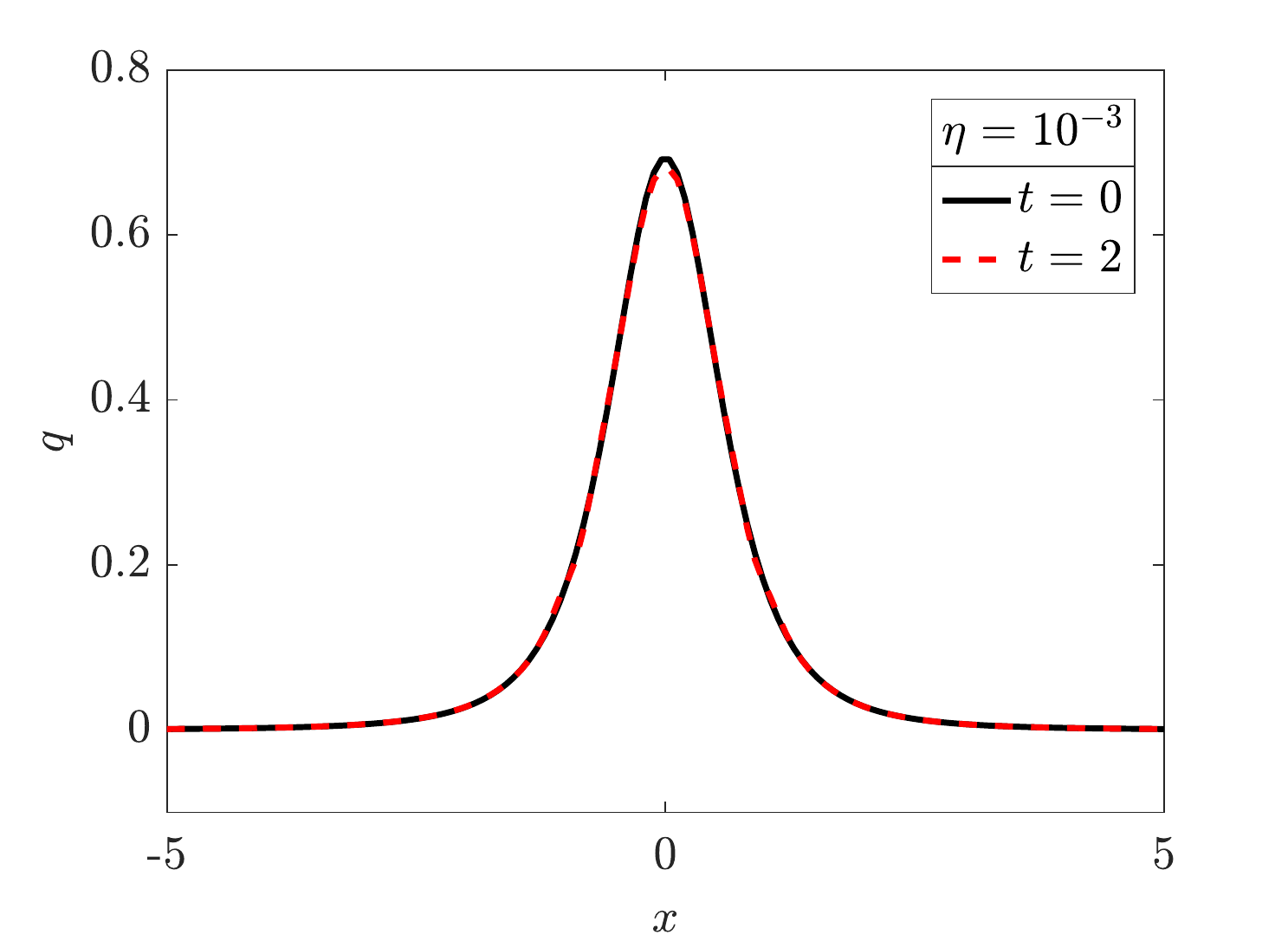}}
\subfloat{\includegraphics[width=0.5\textwidth]{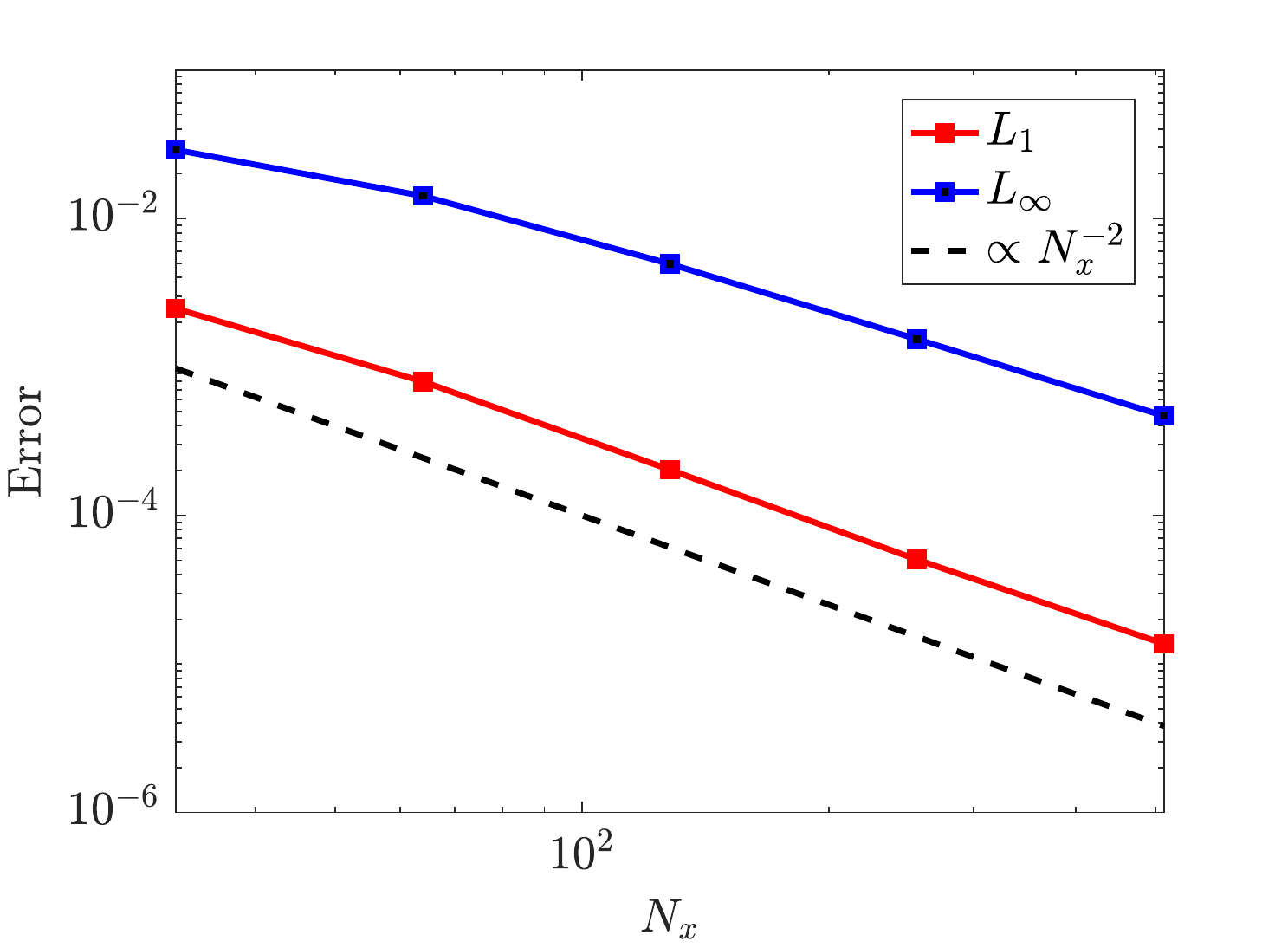}}
\caption{Charged vortex as in \protect\cite{bodomignone}, with in the
  left panel a horizontal ($y=0$) cut of the charge density at $t=0$
  (black line) and $t=2$ (red dashed line) for $256^2$ cells, and in the right panel the
  $L_1$ (red line) and $L_{\infty}$ (blue line) norm of the pressure
  difference between initial and final state versus resolution, showing
  second-order scaling.}
\label{fig:bodomignone}
\end{center}
\end{figure*}

In Fig. \ref{fig:bodomignone} we show a horizontal cut at $y=0$ of the
charge density $q$ at $t=0$ and $t=2$ (left panel), and the second-order
convergence of the $L_1$ and $L_{\infty}$ norms on the difference in
pressure $p$, between the initial and final time, as a function of
resolution (right panel). We point out that our solution is in
  accordance with the results obtained by the application of the CT
  method to control the divergence of both the electric and magnetic
  field variables of \cite{bodomignone}. The large-amplitude
oscillations observed by \cite{bodomignone} in $q$ and $E_y$ and
attributed to a general Lagrange-multiplier method for the divergence
cleaning are absent in our evolution, where we only control the magnetic
field divergence by a CT method instead of both the electric and magnetic
field divergences. Second-order convergence is obtained by maintaining
the equilibrium solution, showing that spatial errors dominate over
temporal errors.

\subsection{Magnetized spherical accretion}
\label{sec:bondi}

\begin{figure*} 
\centering
\includegraphics[width=0.75\columnwidth]{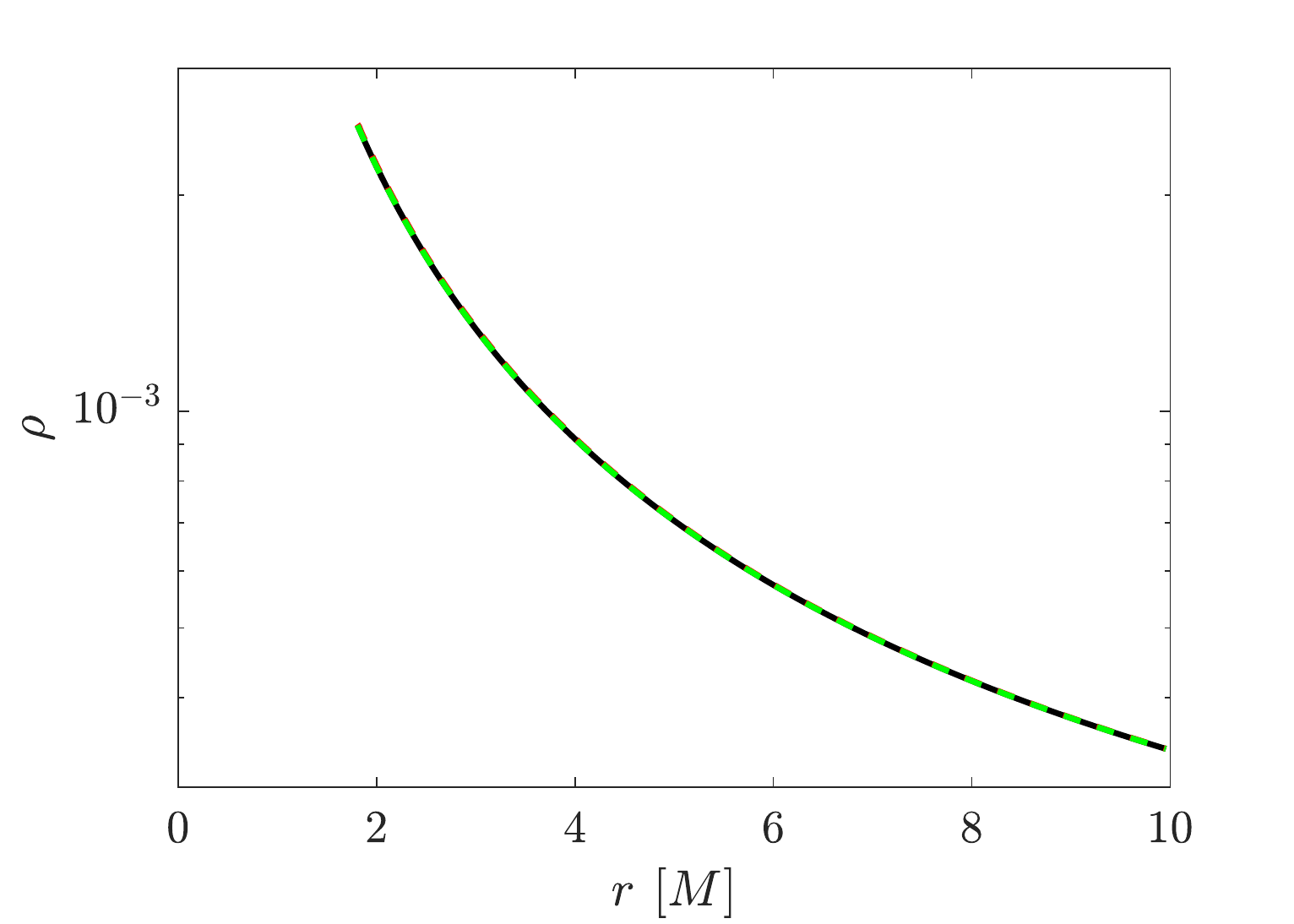}
\includegraphics[width=0.75\columnwidth]{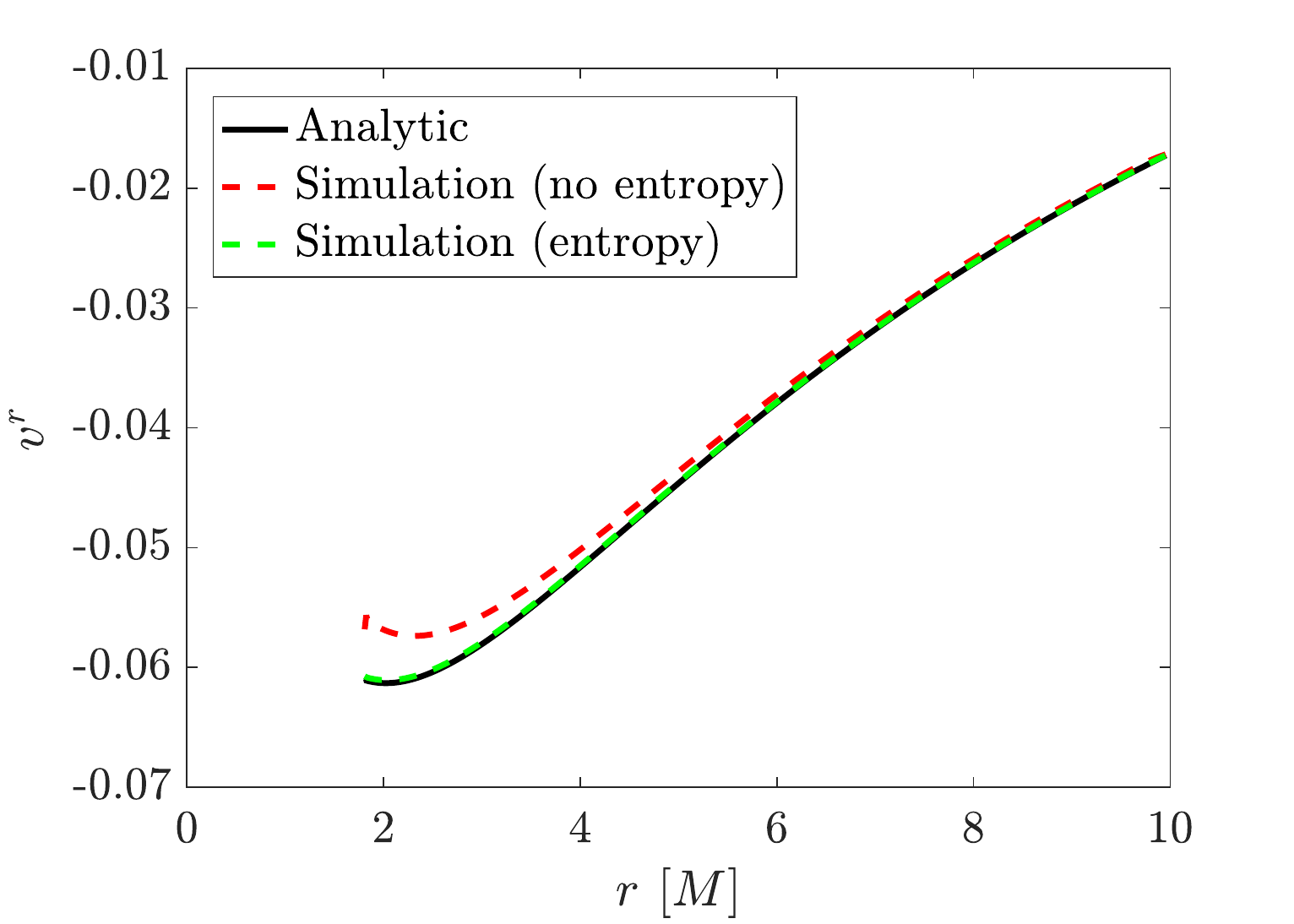}
\includegraphics[width=0.75\columnwidth]{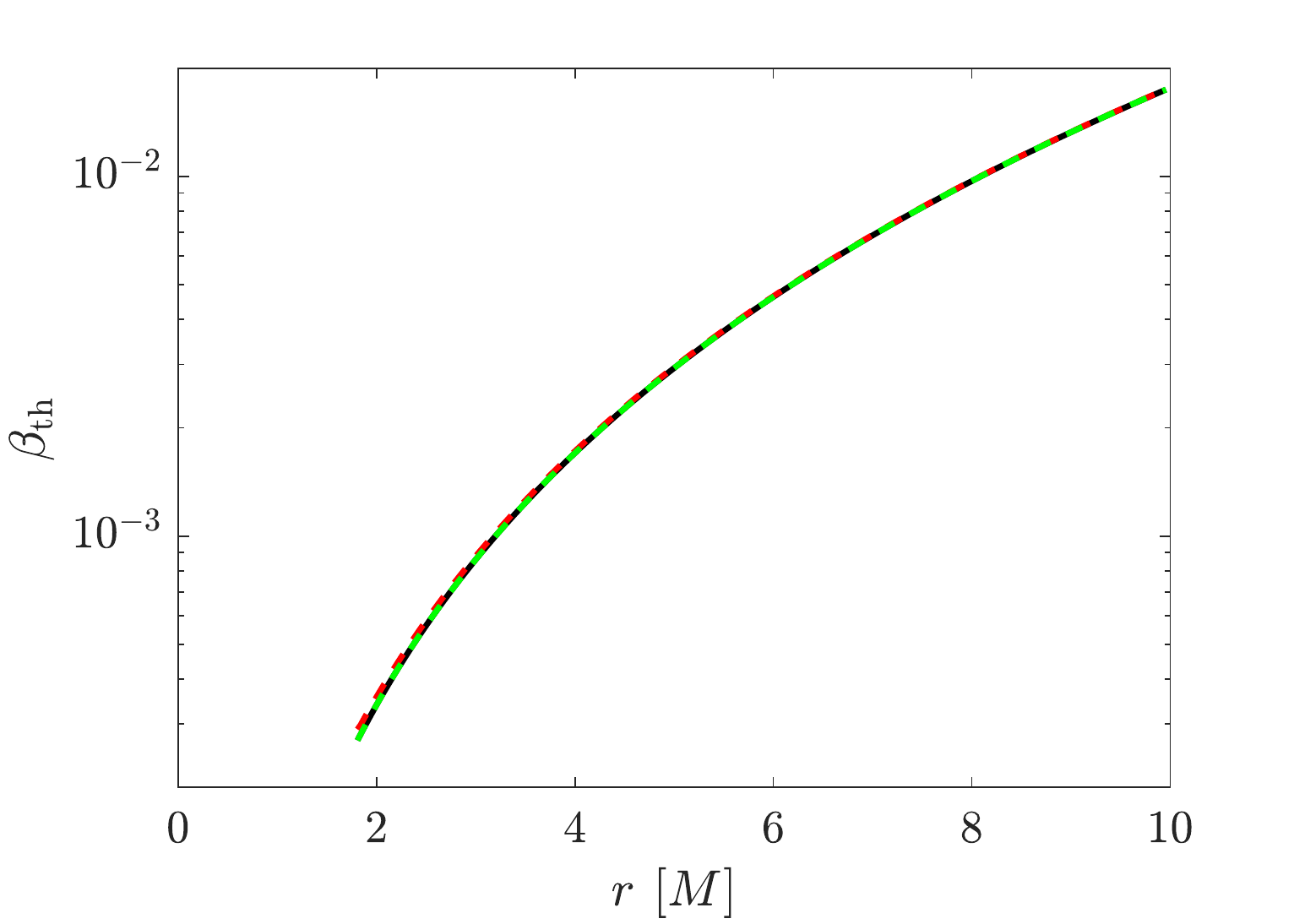}
\includegraphics[width=0.75\columnwidth]{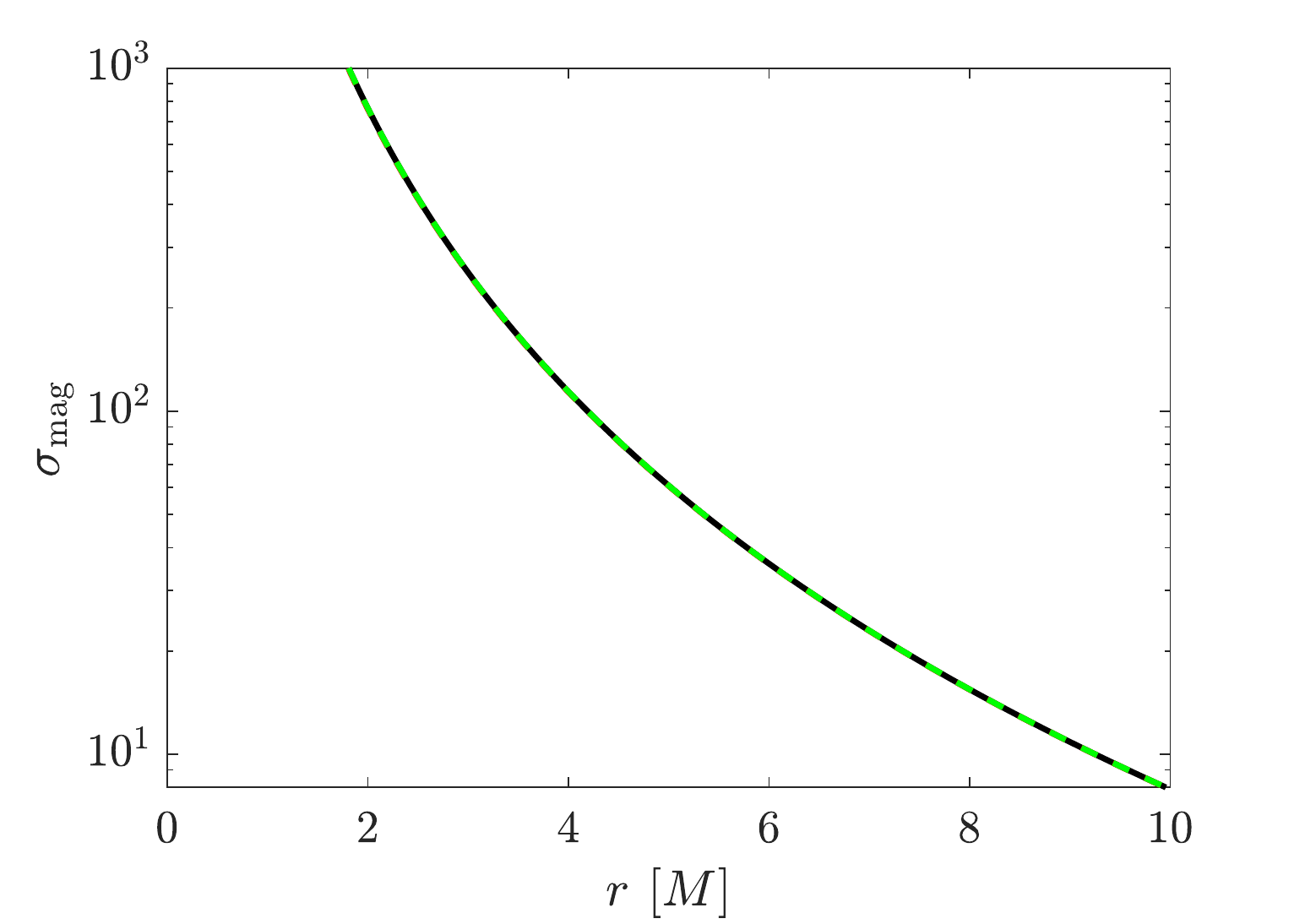}
\caption{Radial profiles of $\rho$, $v^r$, $\beta_\mathrm{th}$, and
  $\sigma_\mathrm{mag}$ for magnetized spherical accretion at $t=100M$ in
  MKS coordinates, with $\sigma_\mathrm{mag} = 10^3$ at the inner edge of
  the domain and a uniform resolution $N_r = 200$. The black solid line 
  indicates the initial exact solution, the dashed red line shows the standard 
  primitive-recovery method, and the dashed green line shows the standard 
  treatment supplemented with the entropy-switch. The entropy-switch is 
  activated for $\beta_\mathrm{th} \leq 10^{-2}$, i.e., for $r \lesssim 8M$. 
  The error in the radial three-velocity $v^r$ shows the clearest advantage of 
  the entropy-switch.}
\label{fig:bondi}
\end{figure*}

As a first test in general relativity, we consider the problem of
spherical accretion onto a Schwarzschild black hole with a strong radial
magnetic field (\citealt{gammie2003}; \citealt{devilliers2003}) and
compare to the steady-state solution (\citealt{bondi1952};
\citealt{michel1972}). This test is a particular challenge for the
primitive-recovery method due to the low $\beta_\mathrm{th}$ regions
close to the event horizon.  We set the mass of the black hole $M=1$ and
dimensionless spin $a = 0$, such that distances and times are measured in
terms of $M$. We follow the initial conditions given by \cite{hawley1984}
and we parametrize the field strength through the magnetization
$\sigma_\mathrm{mag} = 10^3$ at the inner edge of the domain at
$r=1.9M$. To test the inversion method, the resistivity is set to
$\eta=0$, while the full set of GRRMHD equations is solved, allowed by
the first-second IMEX scheme. We employ two-dimensional modified spherical Kerr-Schild
(MKS) coordinates as described in \cite{BHAC} with $r \in [1.9M,10M]; \theta \in [0,\pi]$ and
a uniform radial resolution $N_r = 200$ and angular resolution $N_{\theta} = 100$. The 
steady-state effectively reduces to a one-dimensional problem due to 
the purely radial dependence of the equilibrium solution. 
The analytic solution is fixed at the radial boundaries. The test has been
 reproduced with all 3D and 4D inversion methods. The entropy-switch is 
 activated if $\beta_\mathrm{th} \leq 10^{-2}$ (for $r \lesssim 8M$, see 
 Fig. \ref{fig:bondi}) or if the primary inversion procedure fails.

\begin{figure} 
\centering
\includegraphics[width=1\columnwidth]{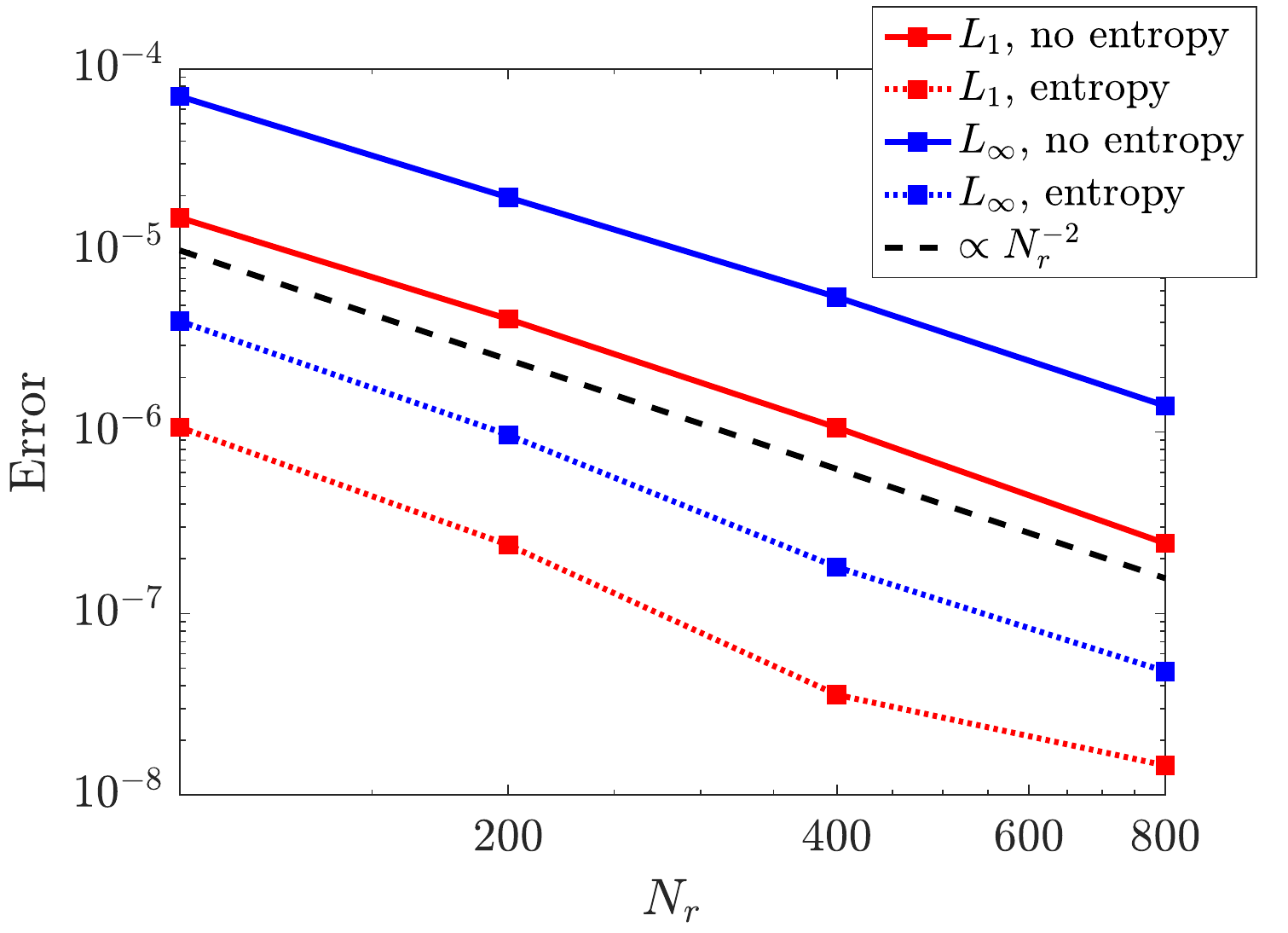}
\caption{Convergence study for the Bondi accretion test at increasing
  number of grid-points $N_r$. The test is run with (dotted lines) and
  without (solid lines) the entropy-switch as a backup strategy for the
  conserved to primitive inversion. The $L_1$ and $L_\infty$ norms of the
  difference between the analytic solution at $t=0$ and the numerical
  result at $t=100$ show second-order convergence in all cases, proving
  that spatial errors dominate over temporal inaccuracies.}
\label{fig:bondiconv}
\end{figure}

Figure \ref{fig:bondi} shows the radial profiles of rest-mass density
$\rho$, radial three-velocity $v^r$, $\beta_\mathrm{th}$, and
$\sigma_\mathrm{mag}$ as found with primitive recovery with entropy
switch (green dashed line) and without (red dashed line) compared to the
analytic solution (black solid line). For the radial three-velocity (top
right panel) it is clear that the entropy-switch improves the solution
close to the event horizon $r \lesssim 6M$. The improvement provided by
the entropy-switch is also visible in the $L_1$ and $L_{\infty}$ norm of
the rest-mass density (Fig. \ref{fig:bondiconv}), increasing the
numerical accuracy by approximately a factor four. Second-order
convergence is obtained both with and without the entropy-switch.

\subsection{Resistive accreting torus}

Finally, we simulate accretion from a magnetized test-fluid
torus (\citealt{fishbone1976}) around a Kerr black hole. We again set the
mass of the black hole $M=1$ and the dimensionless spin $a = 0.9375$. We
employ MKS coordinates on a two-dimensional domain where $r \in [1.29,
  2500]$ and $\theta \in [0,\pi]$ with a uniform resolution of $N_r
\times N_{\theta} = 512 \times 256$ cells. At the initial state, the inner edge of the torus is located
at $r = 6$ and the maximum rest-mass density is localized at $r = 12$. To
simulate the vacuum region outside the torus we set the rest-mass density
and the pressure in the atmosphere as $\rho_\mathrm{atm} =
\rho_\mathrm{min}r^{-3/2}$ and $p_\mathrm{atm} =
p_\mathrm{min}r^{-5/2}$. The rest-mass density and the pressure are reset
whenever they fall below these floor values. The normalization of the
power-law floor model is set to $\rho_\mathrm{min}=10^{-4}$,
$p_\mathrm{min}=(1/3)\times10^{-6}$.

The initial magnetic field configuration consists of a weak single loop
given by the vector potential
\begin{equation}
 A_\phi\propto\mathrm{max}(\rho/\rho_\mathrm{max}-0.2,0),
\end{equation}
where $\rho_\mathrm{max}$ is the global maximum rest-mass density in the
torus. The field strength is determined such that
$2p_\mathrm{max}/b^2_\mathrm{max}=100$, where the spatial locations where
$p_\mathrm{max}$ and $b^2_\mathrm{max}$ are found do not necessarily
coincide. Note that this configuration does not result in an exact MHD
equilibrium.

At the polar axis, we impose
symmetric boundary conditions for all scalar variables, the radial and
poloidal vector components $v^r$, $B^r$, $v^\phi$, $B^\phi$, and the
azimuthal component $E^\theta$; antisymmetric boundary conditions are
imposed for the azimuthal vector components $v^\theta$ and $B^\theta$,
and for the radial and poloidal components $E^r$, $E^\phi$ 
(see \citealt{BHAC} and \citealt{porth2019} for a discussion on the boundary 
conditions in GRMHD simulations of magnetized accretion flows). At the
inner and outer radial boundaries we impose zero-inflow boundary
conditions. We choose an ideal-gas EOS with $\hat{\gamma} = 4/3$.

The initial equilibrium configuration is perturbed with random,
low-amplitude pressure oscillations. This triggers the MRI during the
accretion of gas from the torus onto the central object. The instability
amplifies the initial magnetic field and drives the disruption of the
equilibrium towards a quasi-steady state around $t\sim500 M$. During the
process, the resistivity determines the development of diffusive
processes. Here, we choose a range of values for $\eta \in [10^{-14},
  10^{-2}]$ and we simulate the development of the MRI until $t=2000
M$. We also study the exact ideal-MHD limit $\eta=0$ (allowed by the
first-second IMEX scheme). Additionally, we perform the same simulation
with the ideal-GRMHD version of \texttt{BHAC}. The latter differs from
the $\eta=0$ case simulated with our resistive algorithm in many aspects,
most notably by solving the induction equation for the magnetic field by
assuming that the electric field is a purely dependent quantity $E^i =
-\gamma^{-1/2}\eta^{ijk}v_j {B}_k $ (as presented in
\citealt{BHAC}). Comparing the results of the resistive scheme in the
ideal-MHD limit with the purely ideal-MHD implementation is therefore a
strict and important benchmark. The resistive runs were all completed
within a runtime of a factor $\sim 2-3$ longer compared to the ideal-MHD
case.

\begin{figure*} 
\centering
\subfloat{\includegraphics[height=0.4\textwidth,trim={145mm 0mm 150mm 0mm},clip]{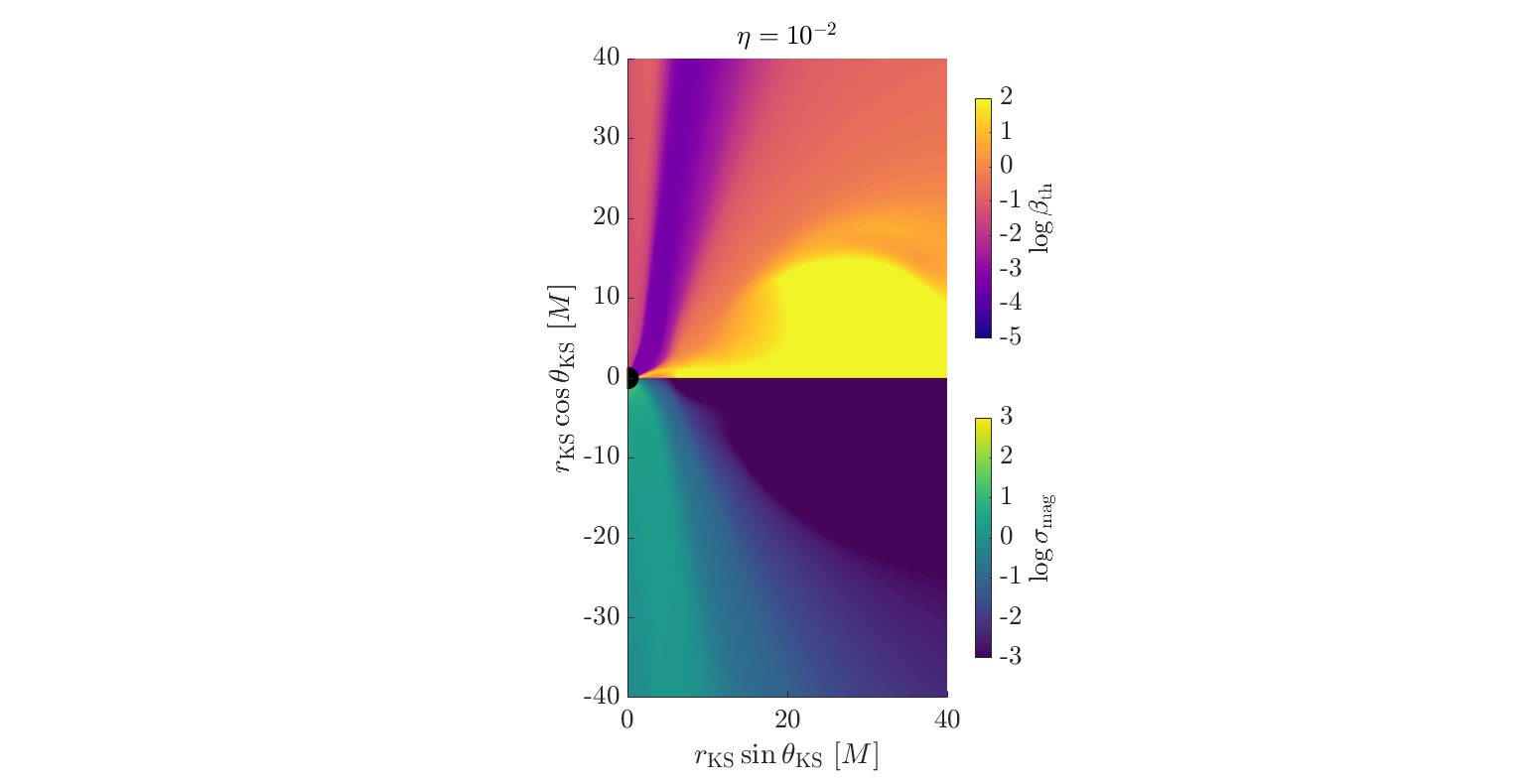}} \qquad
\subfloat{\includegraphics[height=0.4\textwidth,trim={155mm 0mm 150mm 0mm},clip]{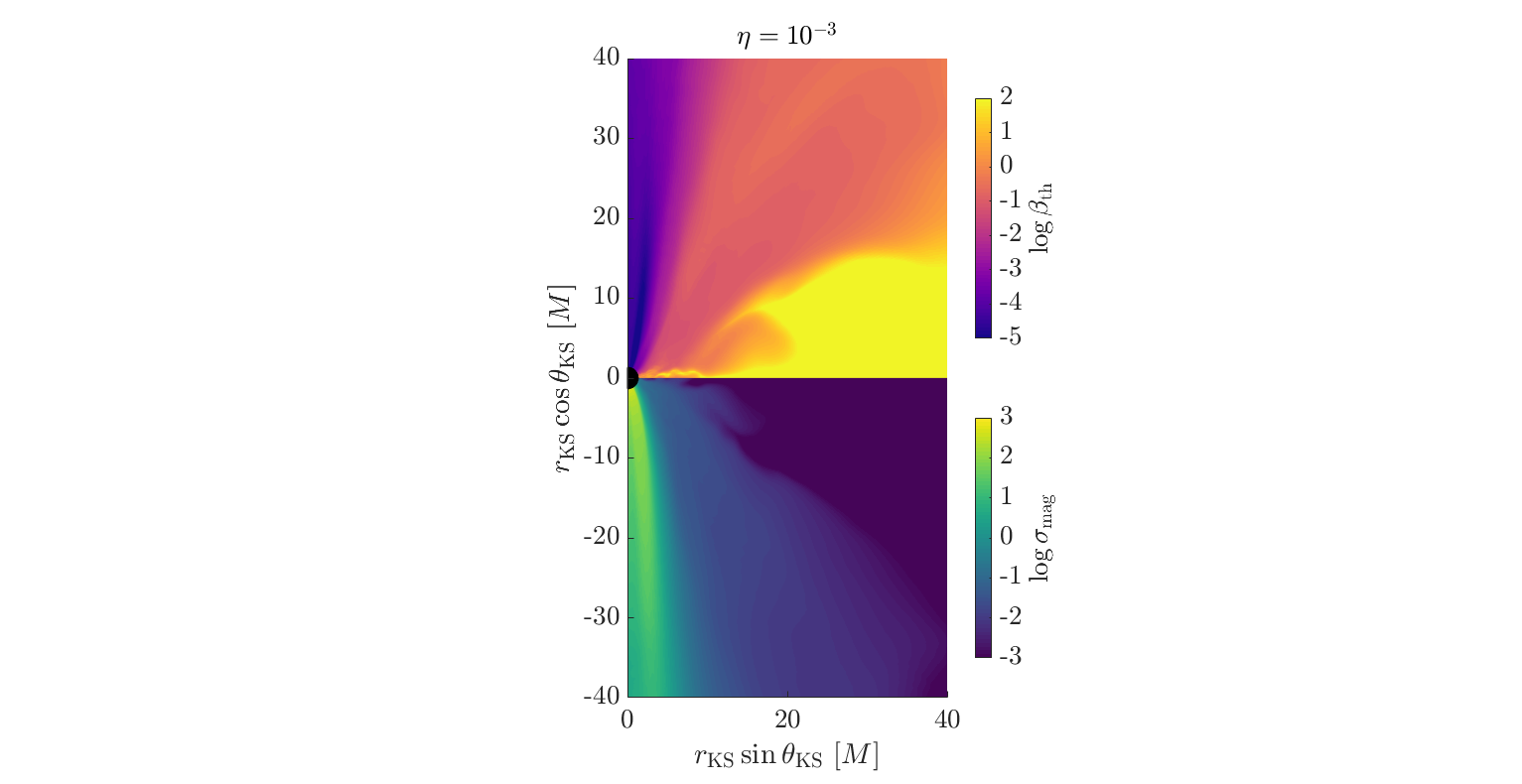}}
\qquad
\subfloat{\includegraphics[height=0.4\textwidth,trim={155mm 0mm 120mm 0mm},clip]{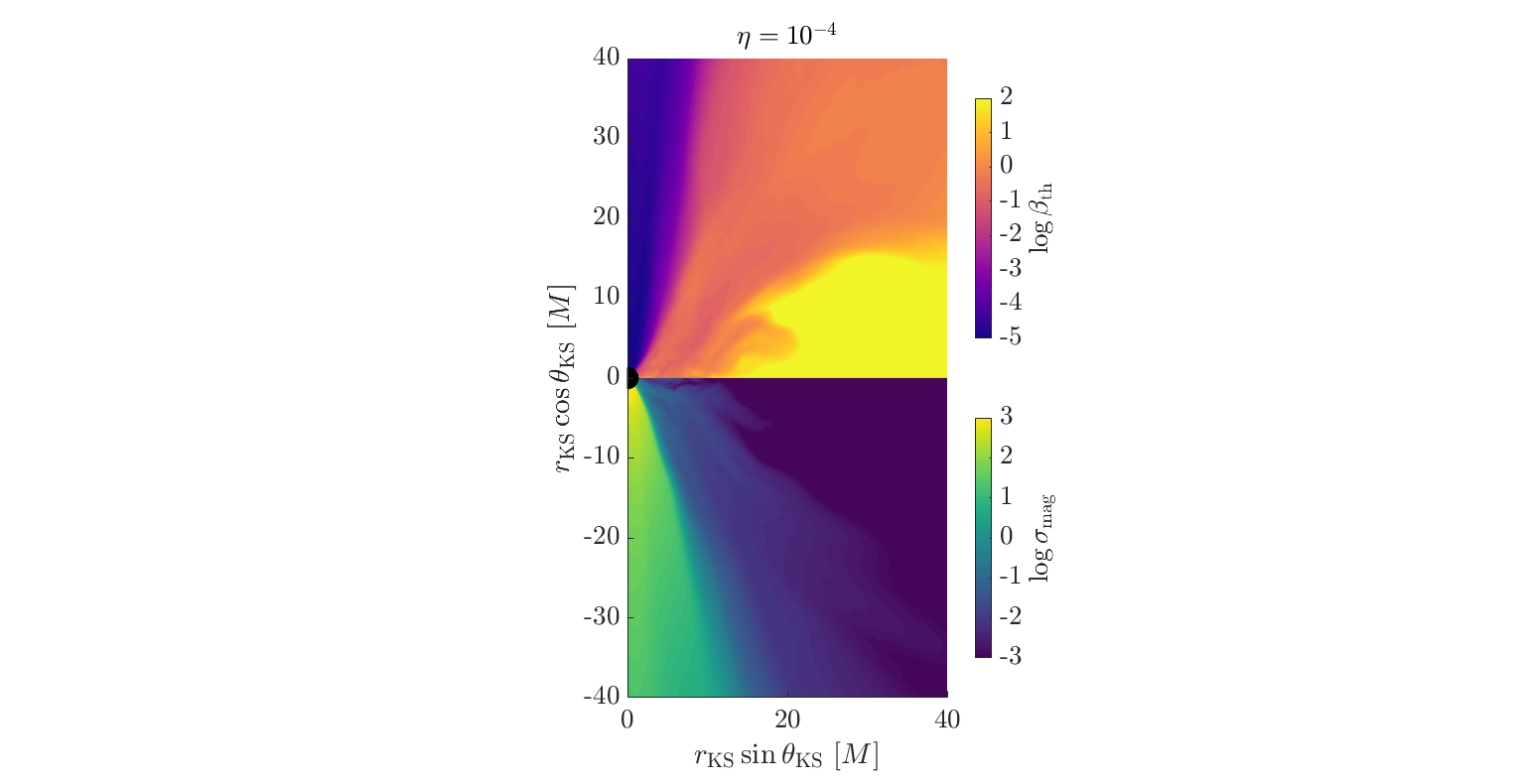}} \\
\subfloat{\includegraphics[height=0.4\textwidth,trim={145mm 0mm 150mm 0mm},clip]{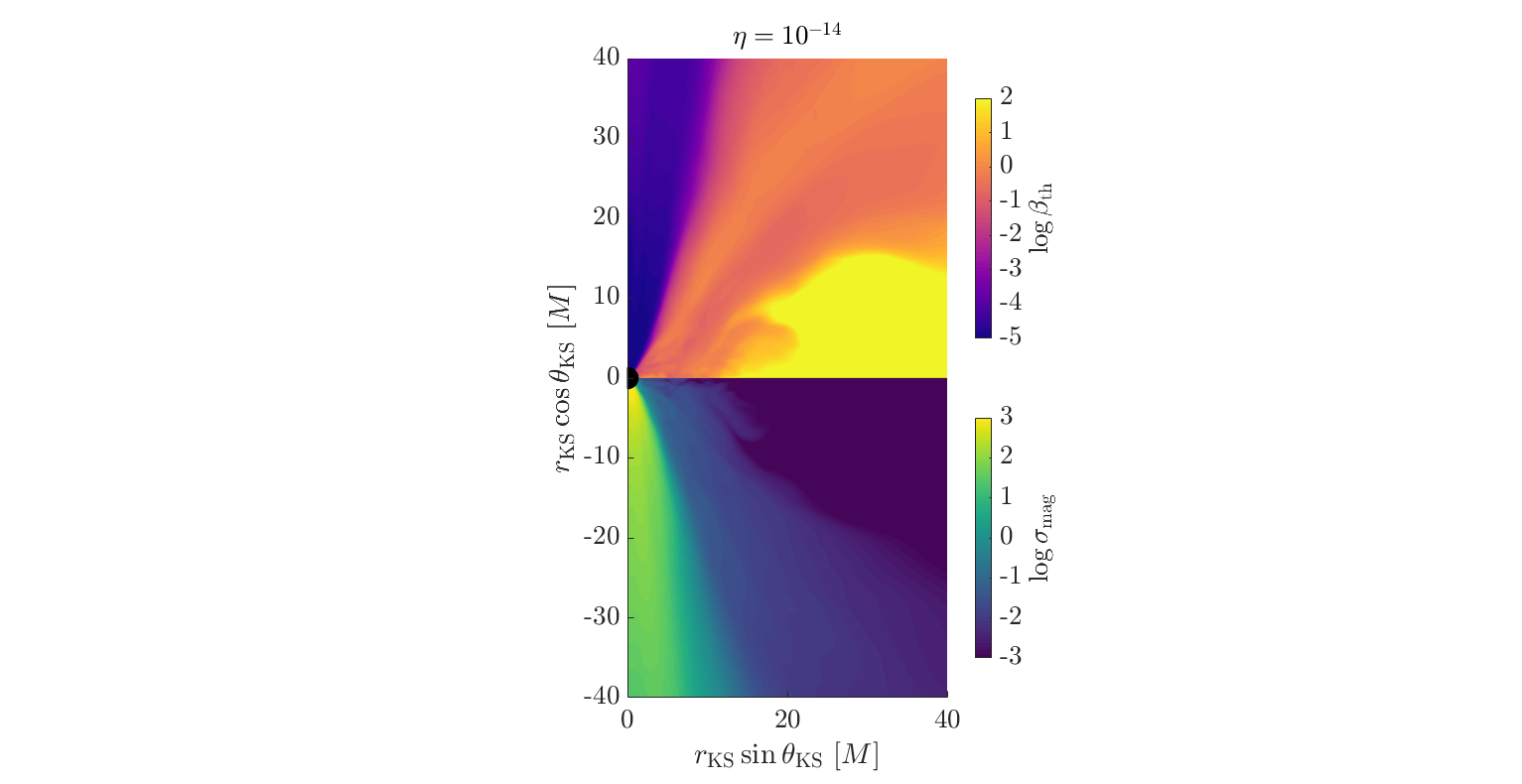}} \qquad
\subfloat{\includegraphics[height=0.4\textwidth,trim={155mm 0mm 150mm 0mm},clip]{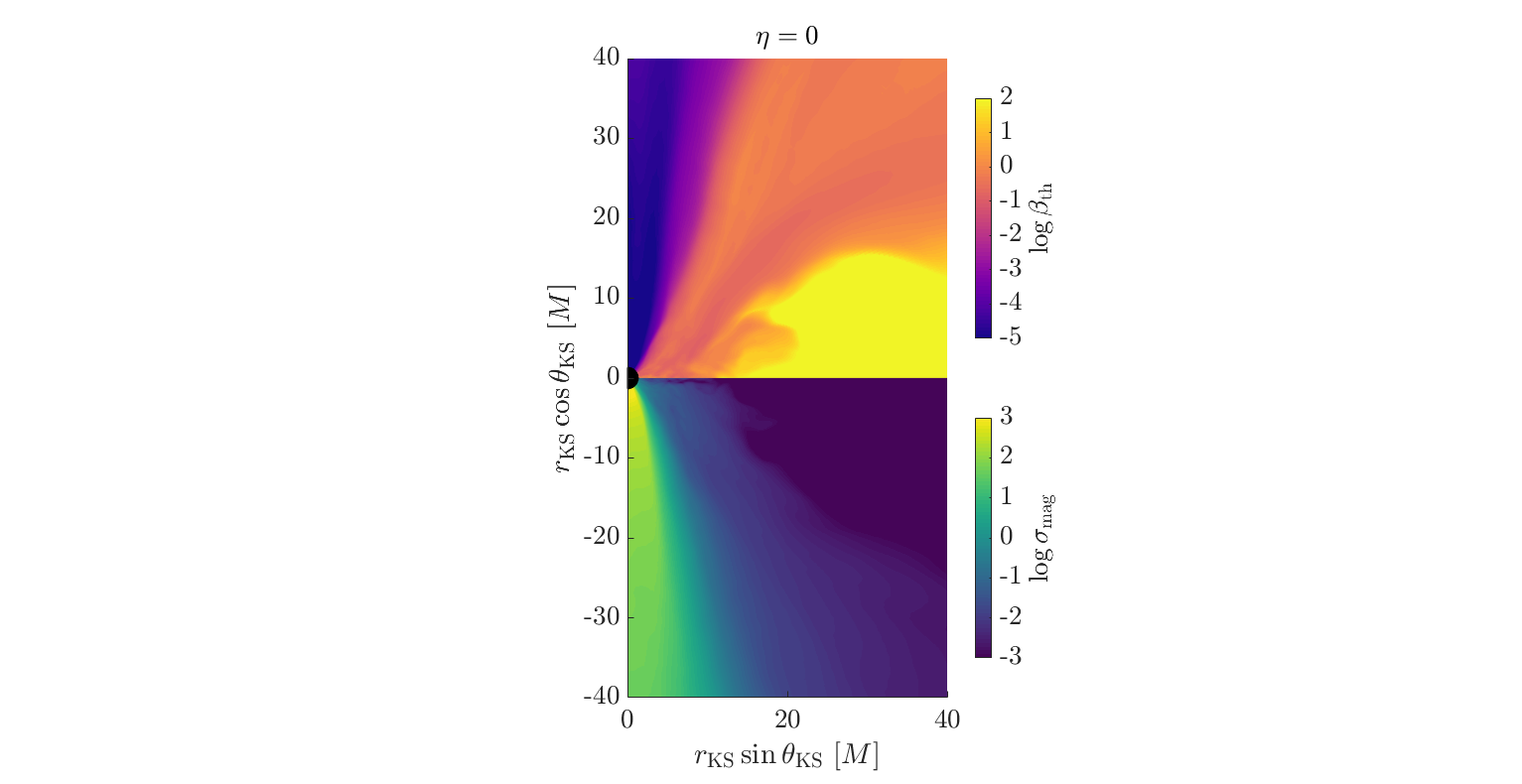}}
\qquad
\subfloat{\includegraphics[height=0.4\textwidth,trim={155mm 0mm 120mm 0mm},clip]{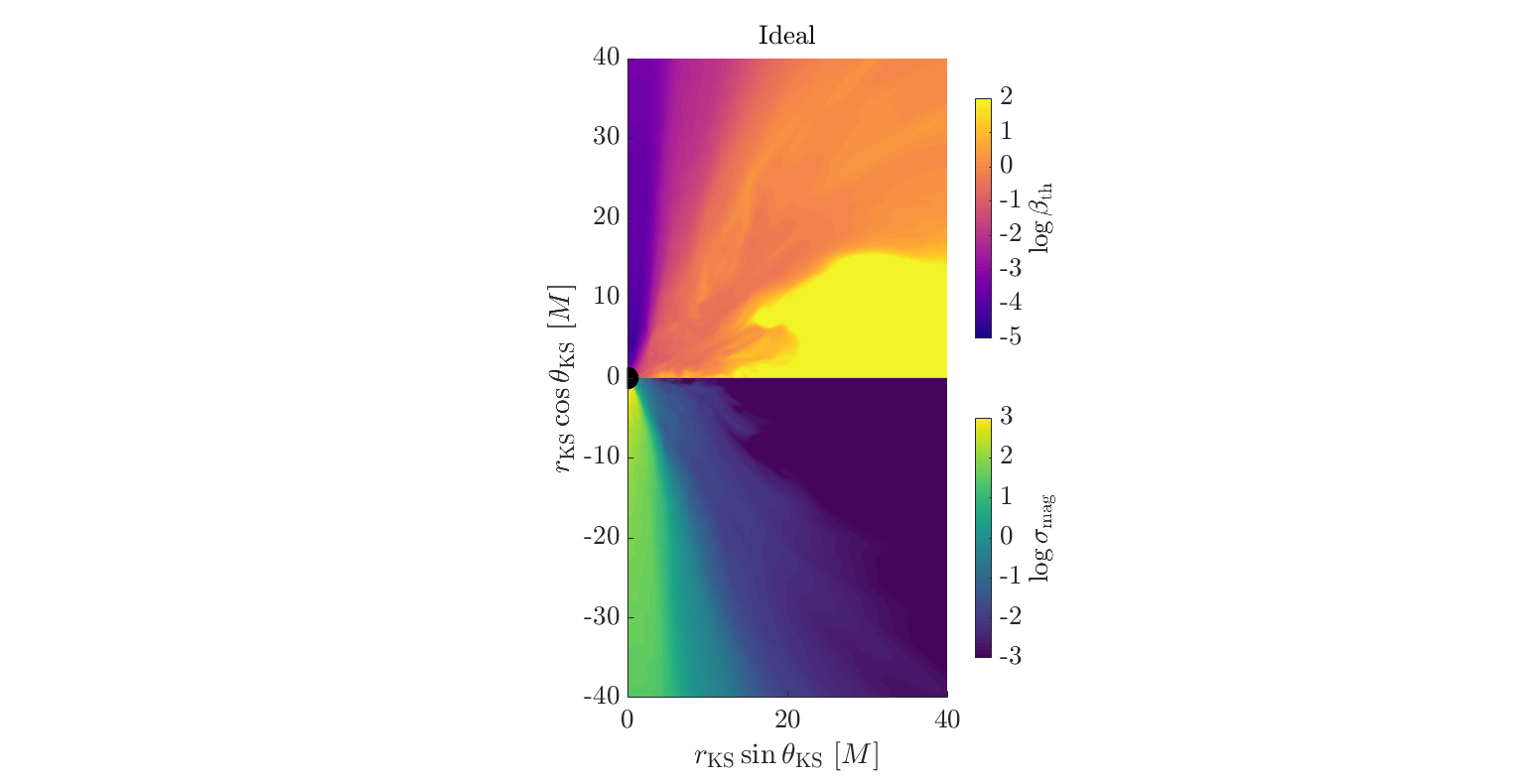}}
\caption{Resistive accreting-torus simulations with
  $\eta=10^{-2},10^{-3},10^{-4}$ (top row, from left to right) and
  $\eta=10^{-14},0$ (bottom row, from left to right) compared to the
  ideal-GRMHD run (bottom right) showing the logarithmic
  $\beta_\mathrm{th}=2p/b^2$ (upper half) and magnetization
  $\sigma_\mathrm{mag}=b^2/\rho$ (lower half) averaged over
  $t\in[500M,1000M]$. The higher resistivity runs show significant
  diffusion and suppression of turbulent structures in the accretion
  flow. The results for lower resistivity $\eta \leq 10^{-4}$ are
  statistically similar, confirming that the numerical resistivity is of
  the order $\eta \sim 10^{-4}$ for the considered resolution, hence
  playing little to no role in the evolution of the system.}
\label{fig:restorus}
\end{figure*}

In Fig. \ref{fig:restorus} we show the spatial distribution of the
characteristic quantities $\beta_\mathrm{th}$ and $\sigma_\mathrm{mag}$
for progressively decreasing $\eta$, and for the simulation run with the
ideal-GRMHD version of \texttt{BHAC}. The results are averaged in time
between $t=500\,M$ and $t=1000\,M$ to account for statistical
fluctuations in the quasi-steady state accretion. The $\eta=10^{-2}$ case
most evidently shows the diffusive effects of resistivity, quenching the
MRI-induced turbulence. 
Turbulent features become progressively more
apparent as $\eta$ decreases, until the point where numerical resistivity
dominates over the explicit physical resistivity. For the resolution
considered here, this threshold can be identified around
$\eta\sim10^{-4}$, at which point the results become visually
indistinguishable from simulations at lower resistivity values (including
the $\eta=0$ limit). Minor visual differences between the ideal-GRMHD
result of \texttt{BHAC} and the $\eta \leq 10^{-4}$ cases are attributed
to the differences in the numerical scheme in the two cases (e.g., the
different characteristic speed employed, see Sec.  \ref{sect:charspeed}).

\begin{figure*} 
\centering
\subfloat{\includegraphics[height=0.4\textwidth,trim={130mm 0mm 130mm 0mm},clip]{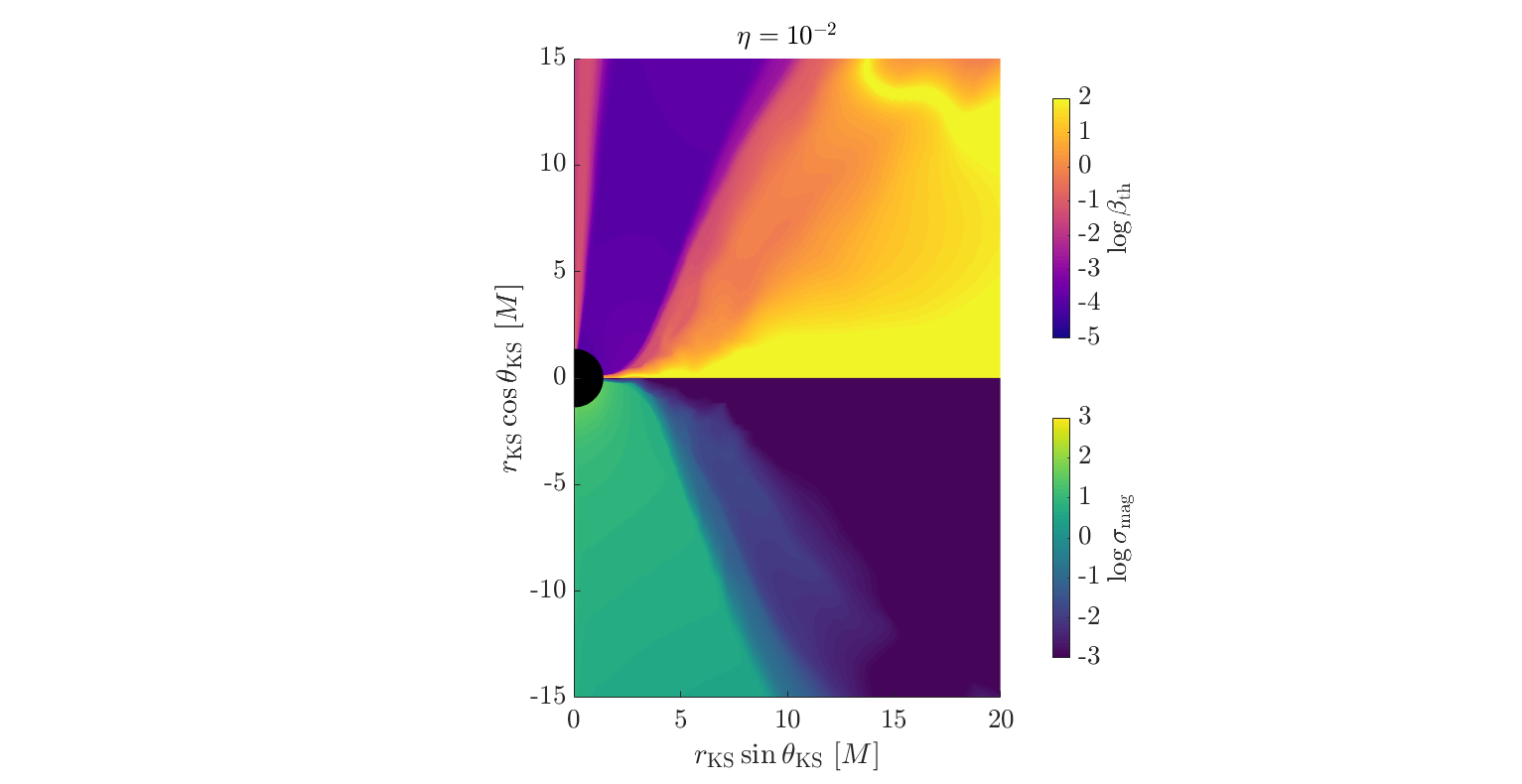}} \qquad
\subfloat{\includegraphics[height=0.4\textwidth,trim={140mm 0mm 100mm 0mm},clip]{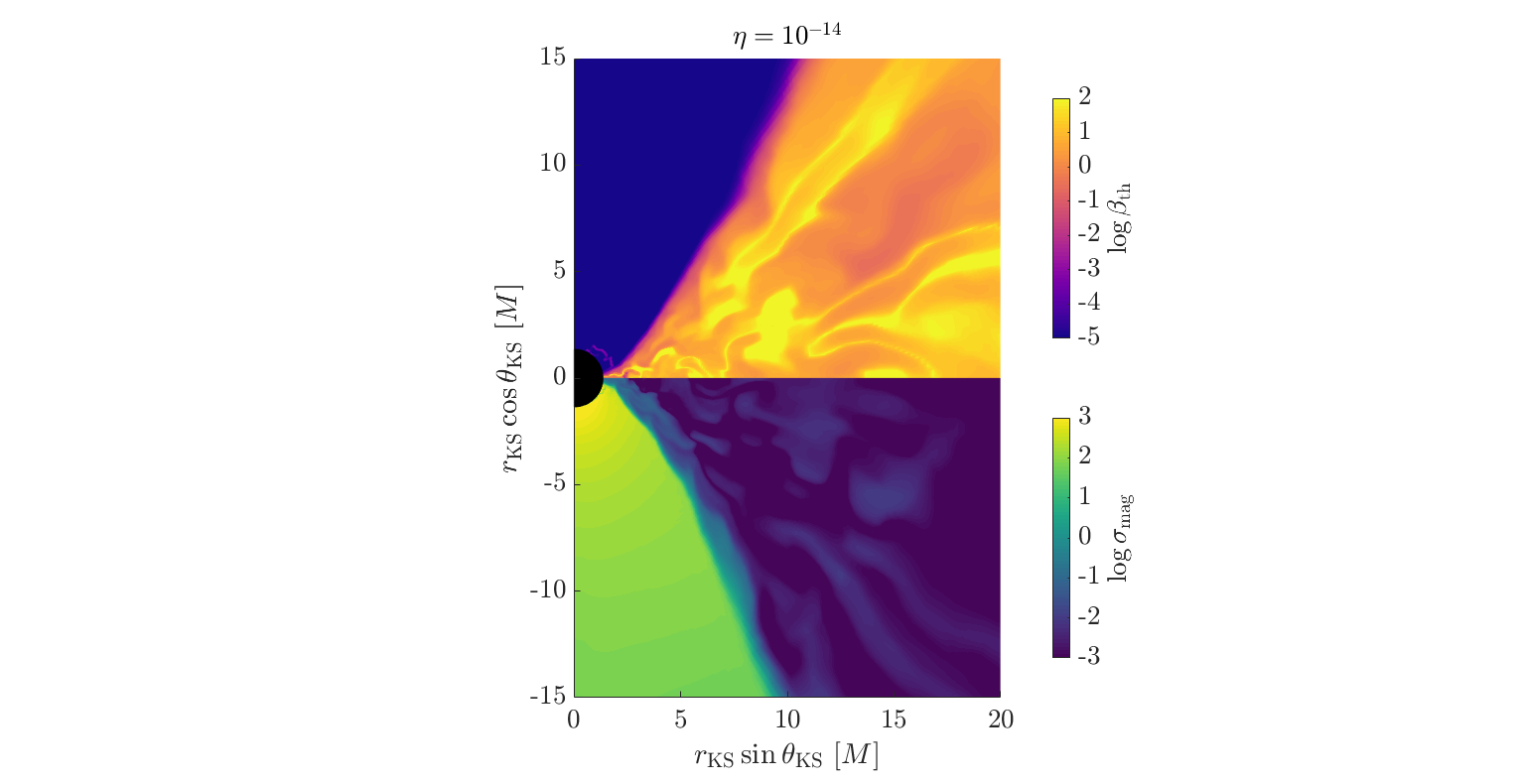}}
\caption{Close-up view of the accretion region in the resistive
  torus simulations with $\eta=10^{-2}$ (left) and
  $\eta=10^{-14}$ (right), showing the logarithmic
  $\beta_\mathrm{th}=2p/b^2$ (upper half) and magnetization
  $\sigma_\mathrm{mag}=b^2/\rho$ (lower half) at time $t=1600 M$. The
  high-resistivity run shows almost no sign of turbulent structures,
  which are instead clearly visible in the low-resistivity case.}
\label{fig:restorus_close}
\end{figure*}

To remove the smoothing introduced by the time averaging,
Fig. \ref{fig:restorus_close} shows a close-up view of the accretion
region at time $t=1600 M$ during the evolution of the system. The
$\eta=10^{-2}$ run (left) shows no sign of turbulence, which is almost
completely suppressed by the diffusive processes introduced by the high
resistivity. The $\eta=10^{-14}$ case (right), on the contrary, clearly
shows the formation of characteristically turbulent structures, with
steep gradients both in $\beta_\mathrm{th}$ and $\sigma_\mathrm{mag}$.

\begin{figure} 
\centering
\includegraphics[width=1\columnwidth]{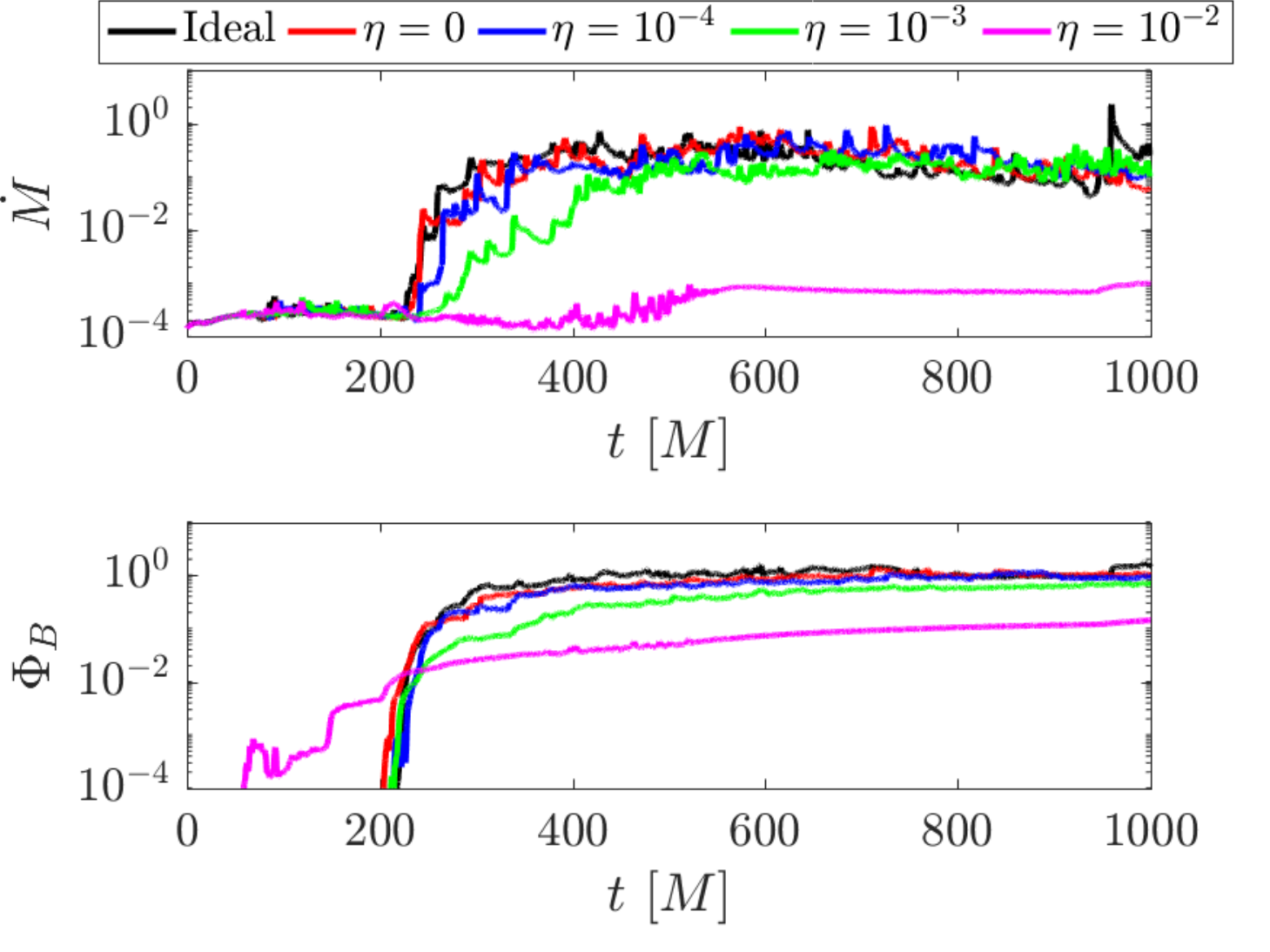}
\caption{Evolution in time of the mass accretion rate $\dot{M}$ (top) and
  magnetic flux through the horizon $\Phi_B$ (bottom) for the resistive
  Fishbone-Moncrief torus with $\eta=10^{-2}$ (magenta lines), $10^{-3}$
  (green lines), and $10^{-4}$ (blue lines). These are compared to the
  $\eta=0$ case (red lines, ideal-MHD limit) and the purely ideal-GRMHD
  run with \texttt{BHAC} (black lines). All cases with $\eta\le10^{-4}$
  show excellent agreement in the MRI development time and steady-state
  values with the ideal-MHD run.}
\label{fig:restorus_mdot}
\end{figure}

Finally, for a more quantitative comparison, we monitor the accretion
rate $\dot{M}$ and magnetic flux through the black hole event horizon
$\Phi_B$ for all runs, defined as
\begin{equation}
 \dot{M}:=\int_0^{2\pi}\int^\pi_0\rho u^r\sqrt{-g}d\theta d\phi,
\end{equation}
\begin{equation}
  \Phi_B:=\frac{1}{2}\int_0^{2\pi}\int^\pi_0 |B^r|\sqrt{-g}d\theta d\phi.
\end{equation}
In Fig. \ref{fig:restorus_mdot} the evolution in time of both quantities
is shown for the $\eta=10^{-4},10^{-3},10^{-2}$ runs, together with the
results from the $\eta=0$ run and the ideal-MHD run. The plots show how
large resistivity, that is above the numerical resistivity threshold
$\eta > 10^{-4}$, can affect the evolution of the system, delaying the
instability in time and decreasing the final semi-steady state
values. These results are consistent with the findings by
\cite{Qian2016}, establishing that a high resistivity significantly
quenches the MRI. Additionally, due to the robust conserved to primitive
strategies presented in Sec. \ref{sect:con2prim}, we find no difficulty
in simulating the demanding cases where $\eta\rightarrow0$. As shown in
Fig. \ref{fig:restorus_mdot}, the development time of the MRI and the
final steady-state values for both $\dot{M}$ and $\Phi_B$ are in good
agreement between the $\eta=0$ run and the ideal-GRMHD simulation. These
are also consistent with the $\eta=10^{-4}$ results, confirming that the
numerical resistivity is of the order of $\eta < 10^{-3}$ for the
resolution considered here. Identifying this threshold is of major
importance, since dissipative length scales that need to be resolved in
resistive simulations are proportional to the resistivity. Hence, the
necessary resolution depends on the resistivity as $N \propto
\eta^{-1}$. If the resolution is lower than the necessary threshold to
capture the resistive dynamics, the numerical resistivity is
prevailing. With explicit resistivity we can explore new physical regimes
that are unattainable in ideal-GRMHD, and explore the effect of
dissipative length scales on the development of the MRI. Explicit
treatment of viscosity in GRMHD (e.g., \citealt{fragile2018};
\citealt{fujibayashi2018}) in combination with resistivity will soon
allow for the investigation of turbulent black hole accretion without
relying on numerical dissipation.

\section{Conclusions}
\label{sect:conclusions}

We presented the implementation of a resistive module in the general
relativistic magneto-fluid code {\tt BHAC}. The new GRRMHD algorithm is
tested and used in this work in combination with AMR and a recently
implemented staggered CT method to ensure solenoidal magnetic fields
(\citealt{olivares2018proc}; \citealt{Olivares2019}).

The GRRMHD equations are solved with the first-second order IMEX scheme
from \cite{Bucciantini} and the performance is compared to the Strang
split scheme of \cite{Komissarov}. The IMEX scheme uses a first-order
iterative implicit step to solve the resistive, stiff term and solves the
non-stiff terms with a second-order explicit scheme as in the ideal-GRMHD
module in {\tt BHAC}. We find that the time-step in the IMEX scheme does
not depend on the resistivity. This results in a speedup compared to the
Strang-split scheme that is of the order of $1/\eta$. Particularly for
cases with $\eta \lesssim 10^{-4}$, this results in a major speedup, since
for this regime the time-step in the Strang-split scheme is dominated by
the resistive stiff terms. The implemented IMEX scheme can be
straightforwardly extended to higher order. Based on the current
implementation, it is also straightforward to incorporate additional
physics like Hall and dynamo dynamics (see e.g., \citealt{Bucciantini};
\citealt{Palenzuela}; \citealt{bugli2014}). The system of GRRMHD
equations can also be extended to evolve an extra equation for radiation
dynamics, where the IMEX scheme is applied to the stiff terms due to the
optically thick plasma (see e.g., \citealt{zanotti2011};
\citealt{roedig2012}; \citealt{sadowski2013}; \citealt{sadowski2014};
\citealt{mckinney2014}, and \citealt{fuksman2019} in SRRMHD).

Well-established GRRMHD methods struggle with regimes where both $\eta$
and plasma-$\beta_\mathrm{th}$ are small (\citealt{Palenzuela2}; \citealt{Qian2016}), e.g., in highly magnetized
accretion flows and jets in the surroundings of black holes and neutron
stars. Here, the dynamic electric field makes the recovery of primitive
variables, a key part of all GRMHD codes, particularly demanding.  We
designed and presented several novel primitive-recovery methods taking
the nonlinear dependence of the dynamic (resistive) electric field on the
primitive variables fully into account. Compared to existing
primitive-recovery methods for GRRMHD presented by
\cite{Dionysopoulou}, and \cite{Palenzuela}, our methods are very robust
in a large parameter space and can accurately handle nonzero and
non-uniform resistivity ranging from the ideal-MHD limit $\eta
\rightarrow 0$ to the electrovacuum limit $\eta \rightarrow \infty$ in
highly magnetized regions of high $\sigma_\mathrm{mag}$ and low
$\beta_\mathrm{th}$. The exact ideal-MHD limit $\eta=0$ is recovered for
several analytic tests and for a realistic accreting torus simulation,
due to the nature of the first-second order IMEX scheme of
\cite{Bucciantini}. We note that the 3D
primitive-recovery method by \cite{Bucciantini} and \cite{bodomignone} performs
similarly well in the tests presented in this work.

Additionally, we proposed a backup system of recovery methods, combined
with an entropy-switch. This combined method turned out to be essential
to accurately resolve highly magnetized regions in black hole accretion
simulations. We explored a parameter space of $\sigma_\mathrm{mag} \in
[10^{-2},10^2]$, $\beta_\mathrm{th}\in [10^{-10}, 10^5]$, $\Gamma-1\in
[10^{-2}, 10^3]$ and $\eta \in [10^{-14}, 10^6]$, which should
representative for all regions normally encountered in simulating
high-energy astrophysical phenomena. Combined with the AMR strategy in
{\tt BHAC}, the new GRRMHD algorithm allows for resolving both the global
accretion features governed by the MRI-induced turbulence, and the
dissipative reconnection physics that are conjectured to be responsible
for non-thermal radiation. These non-thermal processes can be
subsequently modeled in \texttt{BHAC} with first-principle approaches,
e.g., the newly implemented general-relativistic (charged) particle
module (\citealt{ripperda2017}; \citealt{bacchini2018};
\citealt{bacchini2019}).

\section*{Acknowledgements}
This research was supported by projects GOA/2015-014 (2014-2018 KU
Leuven) and the Interuniversity Attraction Poles Programme by the Belgian
Science Policy Office (IAP P7/08 CHARM). BR, FB, OP, and HO are supported
by the ERC synergy grant `BlackHoleCam: Imaging the Event Horizon of
Black Holes' (Grant No. 610058). BR and AN are supported by an Alexander
von Humboldt Fellowship. JT acknowledges support by postdoctoral
fellowship 12Q6117N from Research Foundation -- Flanders (FWO). The
computational resources and services used in this work were provided by
the VSC (Flemish Supercomputer Center), funded by the Research Foundation
Flanders (FWO) and the Flemish Government - department EWI, and by the Iboga cluster at the ITP Frankfurt. BR would like to thank Luca del Zanna, Scott Noble, and Christian
Fendt for sharing details on their codes ECHO and rHARM and Jordy Davelaar, Sasha Philippov, and Lorenzo Sironi for useful discussions and suggestions.

\software{BHAC (\citealt{BHAC}; \citealt{Olivares2019})}



\bibliography{mylib3} 
\bibliographystyle{aasjournal}










\end{document}